\documentclass[aps,prb,twocolumn,superscriptaddress,floatfix,showpacs]{revtex4}
\usepackage{graphicx}
\usepackage{amssymb}
\usepackage{MnSymbol}
\usepackage{amsmath}
\usepackage{verbatim}

\newcommand{\A}{\hat{a}}
\newcommand{\B}{\hat{b}}

\newcommand{\Oo}{\hat{O}}
\newcommand{\Ooi}{O}
\newcommand{\memproj}{\hat{p}}
\newcommand{\sysproj}{\hat{P}}
\newcommand{\sysproji}{P}
\newcommand{\eigQ}{Q}
\newcommand{\Q}{\hat{Q}}
\newcommand{\Qi}{Q}
\newcommand{\q}{\hat{q}}
\newcommand{\vv}{\hat{v}}

\ifx\pdftexversion\undefined
\usepackage[dvips]{hyperref}
\else
\usepackage{hyperref}
\fi
\hypersetup{colorlinks = true, linkcolor = blue}

\begin{document}
\title{Time correlators from deferred measurements}

\author{D.\ Oehri}
\affiliation{Theoretische Physik, ETH Zurich, CH-8093
Zurich, Switzerland}
\author{A.V.\ Lebedev}
\affiliation{Theoretische Physik, ETH Zurich, CH-8093
Zurich, Switzerland}
\author{G.B.\ Lesovik}
\affiliation{L.D.\ Landau Institute for Theoretical Physics, RAS,
142432 Chernogolovka, Russia}
\author{G.\ Blatter}
\affiliation{Theoretische Physik, ETH Zurich, CH-8093
Zurich, Switzerland}

\date{\today}

\begin{abstract}
Repeated measurements as typically occurring in two- or multi-time correlators
rely on von Neumann's projection postulate, telling how to restart the system
after an intermediate measurement. We invoke the principle of deferred
measurement to describe an alternative procedure where co-evolving quantum
memories extract system information through entanglement, combined with a
final readout of the memories described by Born's rule.  The new approach to
repeated quantum measurements respects the unitary evolution of quantum
mechanics during intermediate times, unifies the treatment of strong and weak
measurements, and reproduces the projected and (anti-) symmetrized correlators
in the two limits. As an illustration, we apply our formalism to the
calculation of the electron charge correlator in a mesoscopic physics setting,
where single electron pulses assume the role of flying memory qubits. We
propose an experimental setup which reduces the measurement of the time
correlator to the measurement of currents and noise, exploiting the (pulsed)
injection of electrons to cope with the challenge of performing short-time
measurements.
\end{abstract}


\maketitle

\section{Introduction}

Within the quantum world, the question what quantities can be measured in an
experiment is often a non-trivial one, e.g., measuring time correlators (with
times $\tau_j$) requires finding the correct ordering of operators.  Concrete
examples in mesoscopic physics and quantum optics are the measurements of
charge correlators \cite{lesovik:97,lesovik:98,aguado:00,bayandin:08} or full
counting statistics \cite{levitov:96,bagrets:03,dilorenzo:06} and that of
photon correlators \cite{glauber:63,mandel:66}. The question is usually
resolved by including the measurement apparatus in the description and its
internal workings decide upon the form of the measured correlator. Examples
are the Amp\`eremeter, double-dot detector, and spin counter used in Refs.\
\onlinecite{lesovik:97}, \onlinecite{aguado:00}, and \onlinecite{levitov:96},
or the different photodetectors introduced by Glauber and by Mandel
\cite{mandel-wolf}. These detectors then act back on the system, thereby
influencing the measurement outcome, i.e., the specific form of the
correlator.  E.g., a weak measurement as used in Ref.~\onlinecite{lesovik:97},
see also Ref.\ \onlinecite{belzig:13}, leads to symmetrized (${\cal R}
S(\tau_1, \tau_2)$) and antisymmetrized (${\cal I} S(\tau_1, \tau_2)$)
correlators weighted with different detector response functions
\cite{lesovik:98,sadovskyy:11}, while a strong measurement produces a
projected correlator $S^P(\tau_1,\tau_2)$.

These different forms of measured correlators can be derived \cite{oehri:14}
by invoking the von Neumann projection postulate \cite{neumann:31}, telling
how to restart the system after the first measurement at $\tau_1$ (or after
previous measurements at times $\tau_j < \tau_n$ in a $n$-th order
correlator).  The measurement with a weakly coupled detector can be treated
perturbatively, with the von Neumann projection excerted on the detector and
no back action on the system \cite{lesovik:97}.  In a strong measurement with
a large system--detector coupling, the projection formally can be applied
directly to the system, therefore producing a maximal back action
\cite{oehri:14}.
\begin{figure}
\begin{center}
\includegraphics[width=8cm]{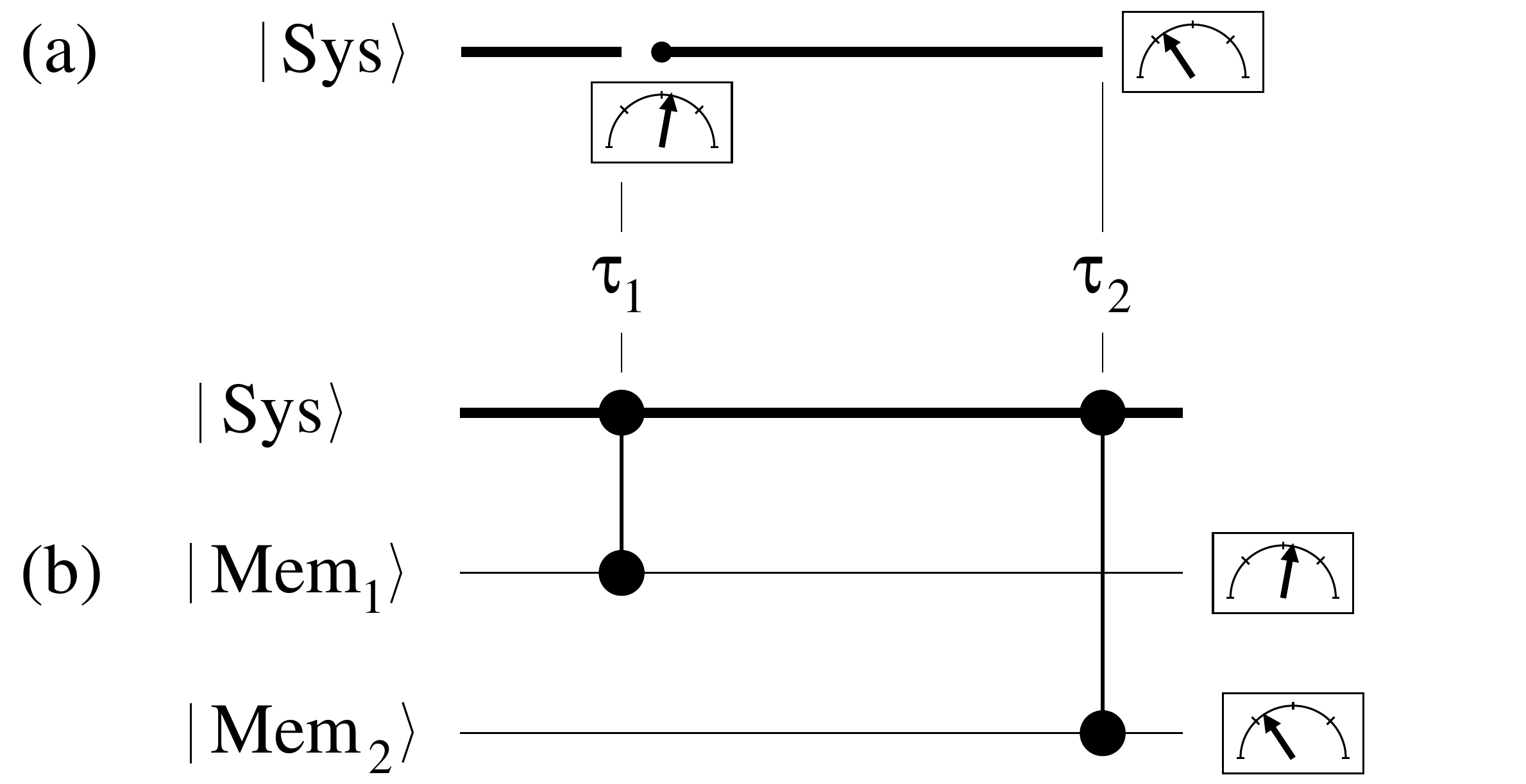}
\end{center}
\caption{\label{fig:comp} Schematic illustration of different measurement
procedures for a two-time ($\tau_1$, $\tau_2 > \tau_1$) correlator: (a) strong
measurement described by von Neumann projection acting directly on the system
at time $\tau_1$ and providing a projected correlator $S^P$ after readout at
$\tau_2$; for a weak measurement, the von Neumann projection at time $\tau_1$
acts on the weakly coupled detector. (b) Repeated measurement without von
Neumann projection at $\tau_1$: unitary co-evolution of system and quantum
memories which are entangled at times $\tau_1$ and $\tau_2$ and final readout
after $\tau_2$.  The coupling strength between the system and the quantum
memories determines the degree of entanglement.}
\end{figure}

In this paper, we invoke the principle of deferred measurement
\cite{nielsen:00} known from quantum information theory, where it can be used
for quantum computing, and apply it to the problem of repeated measurements,
specifically, of time-correlators. We replace the von Neumann projection by
entangling the measured system with co-evolving quantum memories, see Fig.\
\ref{fig:comp}, thereby (effectively) expanding the Hilbert space of the total
system in every measurement step. The desired correlator then is derived from
a final measurement of all the quantum memories by invoking
Born's rule \cite{born:26}. Hence the entire system plus memories undergoes a
unitary quantum evolution until the very end, where the Born rule takes us from the
quantum to the classical world.  The new scheme captures the cases of weak and
strong measurement within a unique formalism by merely changing the degree of
entanglement between the system and the quantum memory.  In the limits of weak
and strong entanglement, we reproduce the results previously derived via use
of the projection postulate.  No simple physical form for the time-correlators
could be found so far in the intermediate coupling regime.

Describing a measurement by entangling the system with a detector and
including a (dissipative) bath in the evolution of the density matrix is a
concept that has been well developed over the past two decades
\cite{gurvitz:97,makhlin:00,averin:05,wiseman:10}. Here, we extend the idea of
system--detector entanglement to the case of repeated measurement. Thereby,
the quantum memories evolve coherently during and after the information
transfer from the system due to entanglement, with the (dissipative)
measurement deferred to the very end of the process.

The proposed scheme for the measurement of time-correlated observables finds
an interesting application in mesoscopic physics. In particular, measuring the
charge $\Q$ dynamics of a quantum dot with the help of a nearby quantum point
contact is a classic problem by now\cite{classic_qpc_qd}. In such a setup,
single electron pulses \cite{levitov:96,feve:07} assume the role of flying
qubit memories which are either transmitted or reflected by the quantum point
contact (QPC), depending on the dot's charge state, see Fig.\
\ref{fig:meso_setup}. Analyzing the charge transmitted across the QPC then
provides the desired information on the dot's charge correlator.  Making use
of recent developments in electron quantum optics
\cite{fletcher:13,bocquillon:13}, we propose a setup, see Fig.\ \ref{fig:exp},
that shifts the task of resolving short times, typically done on the level of
the detection, to the proper timing of electron pulses and gate operations.

We briefly sketch the main idea of the paper: Consider a system observable
$\Oo$ with an eigenbasis $\{|n\rangle\}$ that is to be measured. We
start with a system state at the initial time $\tau_\mathrm{in}$
\begin{align}\label{eq:Psi}
   |\psi(\tau_\mathrm{in})\rangle = \sum_n \psi_n(\tau_\mathrm{in}) |n\rangle
\end{align}
and a quantum memory in the initial state
$|\phi_\mathrm{in}^{\scriptscriptstyle (1)}\rangle$ and have them transiently
interact at time $\tau_1$ with the help of an externally controlled
interaction (to be identified with a quantum detector).  The system and memory
then become entangled,
\begin{align}\label{eq:oneen}
   \bigl[\sum_n \psi_n (\tau_1) |n\rangle\bigr]
   |\phi_\mathrm{in}^{\scriptscriptstyle (1)} \rangle \to \sum_n \psi_n(\tau_1)
   |n\rangle |\phi_n^{\scriptscriptstyle (1)}\rangle,
\end{align}
where $|\phi_n^{\scriptscriptstyle (1)}\rangle$ denote memory states after
interaction with the system in state $|n\rangle$ and we assume a negligible
evolution of the system during the time of interaction. Evolving the 
system to the later time $\tau_2$ and entangling it with a second memory, we
obtain the state
\begin{align}
\sum_{m,n} U_{mn} (\tau_{21}) \psi_n (\tau_1) |m\rangle
|\phi_n^{\scriptscriptstyle (1)} \rangle |\phi_m^{\scriptscriptstyle (2)}
\rangle, 
\end{align}
with $U_{mn}(\tau)$ the matrix elements of the system propagator $\hat{U}
(\tau)$ and $\tau_{21} = \tau_2-\tau_1$.  The unitary evolution of the memory
states $|\phi_n^{\scriptscriptstyle (j)}\rangle$ preserves the system
information gained at times $\tau_j$. At time $\tau_\mathrm{fin} > \tau_2$ we
measure the memory observables $\A^{\scriptscriptstyle (1)}$ and
$\A^{\scriptscriptstyle (2)}$ with discrete spectra $a^{\scriptscriptstyle
(1,2)}_{\alpha}$. Making use of Born's rule, we find the probability
distribution function $P_{\alpha\beta}$ for the measurement outcomes $\alpha$
and $\beta$ on the two memory observables; this probability distribution
contains the desired information on the system's two-time correlator.  The
specific relation between the distribution function $P_{\alpha\beta}$ of
measurement outcomes and the correlator of system observables $\Oo(\tau_1)$
and $\Oo(\tau_2)$ depends on the system--detector coupling and the observables
measured on the memories; we will show below how to extract the well known
(anti-)symmetrized and projected correlators from the probabilities
$P_{\alpha\beta}$ in the limits of weak and strong measurements.

In the end, by making use of quantum memories which store the information
acquired from the system at the quantum level at earlier times $\tau_1$ and
$\tau_2$ until the final time $\tau_\mathrm{fin}$, we have avoided the
intermediate readout which requires the use of the projection postulate. Hence
the entire measurement process follows a unitary quantum evolution until the
transition to the classical world is done via Born's rule.

In the following, we will first derive the general framework describing the
deferred measurement of a correlator with the intermediate von Neumann
projection replaced by a system--detector entanglement (Sec.\ \ref{sec:uni}).
In Section \ref{sec:gen_wc}, we discuss the limit of weak measurement and use
qubits as quantum memories to arrive at a simple relation between the
measurement outcome on the qubits and the (anti-) symmetrized correlators of
the system. In Sec.\ \ref{sec:gen_sc}, we first discuss a strong measurement
at strong coupling using qudit memories and then invoke (weakly coupled) qubit
registers to show that both types of strong measurements produce the projected
time correlator of the system. An illustration of our formalism is given in
Sec.\ \ref{sec:charge}, where we describe the measurement of the charge
correlator in a mesoscopic setting, specifically, the two-time charge
correlator of a quantum dot (QD) as measured by a quantum point contact (QPC).
Sec.\ \ref{sec:charge_exp} describes a possible experimental implementation
and in Sec.\ \ref{sec:con} we summarize our results.

\section{Correlator measurements by quantum memories}\label{sec:uni}

We consider the situation where a two-time correlator of a system operator
$\Oo$ is measured with the help of two quantum memories; the system's initial
state $|\psi\rangle$ is given by Eq.\ (\ref{eq:Psi}), while the memories are
described by initial states $|\phi^{\scriptscriptstyle (j)}_\mathrm{in}
\rangle$, $j = 1,2$ (see below for the discussion of an open system described
by a density matrix $\rho$). The memories interact with the system at times
$\tau_{1,2}$ during a small time intervall $\delta \tau$.  After this
interaction, the resulting memory states $|\phi^{\scriptscriptstyle (j)}_{n}
\rangle = \hat{u}_n |\phi^{\scriptscriptstyle (j)}_\mathrm{in} \rangle$ depend
on the system state $|n\rangle$, where $\hat{u}_n$ describes the time
evolution of the quantum memories during the interaction with the system (we
assume a trivial evolution of the free memories). The system state
$|\psi(\tau)\rangle = \sum_n \psi_n(\tau) |n\rangle$ is assumed to remain
unchanged during the time $\delta \tau$ of the individual interaction
events, see Sec.\ \ref{sec:finitewidth} for an extended discussion of this
point.  After the second interaction event at $\tau_2$, the wave function
$|\psi(\tau_\mathrm{fin} > \tau_2) \rangle$ of the system is entangled with
the states $|\phi^{\scriptscriptstyle (j)}_{n} \rangle$ of the memories and
the combined wave function $|\Psi_f\rangle$ reads
\begin{equation}\label{eq:Psif}
   |\Psi_f\rangle = \sum_{l,m,n} U_{lm}(\tau_{f2}) 
                                 U_{mn}(\tau_{21}) \psi_n(\tau_1)\, |l\rangle
      \, |\phi^{\scriptscriptstyle (1)}_{n} \rangle
      \, |\phi^{\scriptscriptstyle (2)}_{m} \rangle,
\end{equation}
with $\tau_{f2} = \tau_\mathrm{fin} - \tau_2$. The quantum memories are
supposed to keep their system information after their interaction.  At time
$\tau_\mathrm{fin}$, we measure the operators $\A^{\scriptscriptstyle
(1)}$ and $\A^{\scriptscriptstyle (2)}$ on the first and second
memory, respectively. Denoting the (discrete) eigenvalues and eigenstates of
$\A^{\scriptscriptstyle (j)}$ by $a_{\alpha}^{\scriptscriptstyle
(j)}$ and $|\varphi_{\alpha}^{\scriptscriptstyle (j)} \rangle$, we rewrite the
memory states $|\phi_{n}^{\scriptscriptstyle (j)} \rangle = \sum_{\alpha}
s_n^{\alpha} |\varphi_{\alpha}^{\scriptscriptstyle (j)} \rangle$.  Applying
Born's rule to the final state (\ref{eq:Psif}) provides us with the
probability distribution
\begin{equation} \label{eq:Pabdef}
   P_{\alpha\beta}(\tau_{21}) =
   \langle \Psi_f| \memproj_\alpha^{\scriptscriptstyle (1)} 
           \memproj_\beta^{\scriptscriptstyle (2)} |\Psi_f \rangle,
\end{equation}
where $\memproj_{\alpha}^{\scriptscriptstyle (j)}$ is the projector onto
the eigenstate $\alpha$ of the $j$-th memory, i.e.,
$\memproj_{\alpha}^{\scriptscriptstyle (j)} =
|\varphi_{\alpha}^{\scriptscriptstyle (j)} \rangle \langle
\varphi_{\alpha}^{\scriptscriptstyle (j)}|$. Making use of the unitarity
condition $\sum_{l} U_{lm} U^\ast_{lm'} = \delta_{mm'}$ (rendering the
evolution $U_{lm}(\tau_{f2})$ in Eq.\ (\ref{eq:Psif}) irrelevant), we obtain
the probabilities
\begin{equation} \label{eq:Pab}
   P_{\alpha\beta}(\tau_{21}) 
   = \sum_m \Bigl| \sum_n s_m^\beta U_{mn}(\tau_{21}) 
                          s_n^\alpha \psi_n(\tau_1) \Bigr|^2.
\end{equation}
The above expressions (\ref{eq:Pabdef}) and (\ref{eq:Pab}) for the
probabilities $P_{\alpha\beta}(\tau_{21})$ provide us with the desired
information on the two-time correlator of the system.  
They are easily generalized to the case of open systems by introducing the
combined system plus bath (the open system) density matrix $\hat{\rho}$ and
evolve it in time including the subsequent entanglement with the quantum
memories: Starting from the initial density matrix $\hat{\rho}_{0} \otimes
|\phi_{{\rm in}}^{\scriptscriptstyle (1)} \rangle\langle \phi_{{\rm
in}}^{\scriptscriptstyle (1)}| \otimes |\phi_{{\rm in}}^{\scriptscriptstyle
(2)} \rangle\langle \phi_{{\rm in}}^{\scriptscriptstyle (2)}|$ describing the
open system plus memories at time $\tau_{\rm in}$, we proceed as in the case
of isolated systems by conditioning the time evolution of the memory states on
the corresponding system states and obtain the final density matrix at time
$\tau_f$
\begin{align}
   \hat{\rho}_f &= \!\!\!\!\!\! \sum_{k,k',n,n',m,m} 
   \!\!\!\!\!\! \hat{U}_{km}(\tau_{f2})
   \hat{U}_{mn}(\tau_{21}) \, \hat{\rho}_{nn'}(\tau_1) \,  
   \hat{U}^\dagger_{n'm'}(\tau_{21}) \label{E:rho_f}\\
   \noalign{\vspace{-2pt}}
   &\qquad\quad
   \times \hat{U}^\dagger_{m'k'}(\tau_{f2})\, |k\rangle \langle k'| \otimes
   |\phi_{n}^{\scriptscriptstyle (1)}\rangle
   \langle \phi_{n'}^{\scriptscriptstyle (1)}| \otimes
   |\phi_{m}^{\scriptscriptstyle (2)}\rangle
   \langle \phi_{m'}^{\scriptscriptstyle (2)}|,\nonumber
\end{align}
with the open system's density matrix $\hat{\rho}(\tau_1)$ at time $\tau_1$,
its reduced part $\hat{\rho}_{nn'}(\tau_1) = \langle n|\hat{\rho}|n'\rangle$,
and the reduced operators $\hat{U}_{il} = \langle i|\hat{U}|l\rangle$ with
$\hat{U}$ the evolution operator of the open system.  Note that here, the
outgoing states $|\phi^{\rm \scriptscriptstyle (1)}_{n^{\rm\scriptscriptstyle
(\prime)}}\rangle$ and $|\phi^{\rm\scriptscriptstyle
(2)}_{m^{\rm\scriptscriptstyle (\prime)}}\rangle$ are conditioned on the
system states $n^{\rm\scriptscriptstyle (\prime)}$ and
$m^{\rm\scriptscriptstyle (\prime)}$ at times $\tau_1$ and $\tau_2$.  We
define the probabilities $P_{\alpha\beta}$ as
\begin{align}\label{eq:Pabdef_o}
   P_{\alpha \beta}(\tau_{21}) &= {\rm Tr}
   \bigl[ \memproj_\alpha^{\scriptscriptstyle (1)}
           \memproj_\beta^{\scriptscriptstyle (2)} \hat{\rho}_f
   \bigr]
\end{align}
with the trace taken over both the open system and the memory states.
Calculating this expression with the final density matrix Eq.\
\eqref{E:rho_f}, we obtain
\begin{align}\label{eq:Pab_o}
   P_{\alpha\beta}(\tau_{21})&=\sum_{n,n',m}  {\rm Tr} \bigl[
   s^{\beta}_{m} \, \hat{U}_{mn}(\tau_{21})\, s^{\alpha}_{n}
   \, \hat{\rho}_{nn'}(\tau_1) \\ 
   \noalign{\vspace{-2pt}}
   &\hspace{80pt}\times
   s^{\alpha \ast}_{n'}\, \hat{U}_{n'm}^\dagger(\tau_{21}) 
   \, {s^{\beta\ast}_{m}}\bigr], \nonumber
\end{align}
with the remaining trace taken over the bath degrees of freedom.  This result
is the direct generalization of \eqref{eq:Pab} to the case of open systems.
The expressions (\ref{eq:Pabdef}) and (\ref{eq:Pab}) as well as
(\ref{eq:Pabdef_o}) and (\ref{eq:Pab_o}) constitute the basic formulas which
we will further develop in the following sections. Indeed, as expressed in
terms of evolution amplitudes of system and detectors, it is difficult to
appreciate the physical meaning and content of these results. In order to make
progress, we consider next the two cases of weak and strong measurements.

\medskip

\subsection{Weak measurement}\label{sec:gen_wc}

Given a {\it weak system--detector coupling}, the most direct way to find the
probabilities $P_{\alpha\beta}$ in terms of physically transparent quantities
is to start from Eq.\ (\ref{eq:Pabdef}) and evaluate this expression
perturbatively in the linear system--detector coupling $H_\mathrm{sd} =
\sum_j\B^{\scriptscriptstyle (j)}(\tau) \Oo$, where the time-dependent
coupling $\B^{\scriptscriptstyle (j)}(\tau)$ acts on the $j$-th memory during
a time $\delta\tau$ around $\tau_j$. The unperturbed evolution of the memories
is described by the Hamiltonian $\hat{h}_0$ and we make use of the interaction
representation. We go over to irreducible quantities by subtracting the
uncorrelated contribution,
\begin{align}\label{eq:Pirr}
  P^\mathrm{irr}_{\alpha\beta} = P_{\alpha\beta} - P_\alpha^{\scriptscriptstyle (1)}
  P_\beta^{\scriptscriptstyle (2)}
\end{align}
with $P_\alpha^{\scriptscriptstyle (1)} = \langle \Psi_f^{\scriptscriptstyle
(1)}| \memproj_\alpha^{\scriptscriptstyle (1)} |\Psi_f^{\scriptscriptstyle
(1)} \rangle$ and $P_\beta^{\scriptscriptstyle (2)} = \langle
\Psi_f^{\scriptscriptstyle (2)}| \memproj_\beta^{\scriptscriptstyle (2)}
|\Psi_f^{\scriptscriptstyle (2)} \rangle$ describing measurements involving
a single entanglement at time $\tau_1$ or $\tau_2$ with only one memory,
respectively.  The quantity $P_\alpha^{\scriptscriptstyle (1)}$ can be
obtained by a simple summation of $P_{\alpha\beta}$,
\begin{align}\label{eq:Pred}
   P_\alpha^{\scriptscriptstyle (1)} = \sum_\beta P_{\alpha\beta},
\end{align}
and $P_\alpha^{\scriptscriptstyle (2)} = P_\alpha^{\scriptscriptstyle (1)}$
for a time-independent problem (otherwise, the determination of
$P_\alpha^{\scriptscriptstyle (2)}$ necessitates a second measurement). Note
that the sum over the first index of $P_{\alpha\beta}$ already includes
correlations, see Eq.\ (\ref{eq:Pab}), and hence $P_\beta^{\scriptscriptstyle
(2)} \neq \sum_\alpha P_{\alpha\beta}$.

The task then is to evaluate the irreducible expression
\begin{eqnarray}\label{eq:Pirrab}
   P^\mathrm{irr}_{\alpha\beta} \! = \! \llangle \Psi |
   \hat{U}_{\scriptscriptstyle D}^\dagger(\tau_f,\tau_{\rm in})
   \memproj_{\alpha}^{\scriptscriptstyle (1)}(\tau_f)
   \memproj_{\beta}^{\scriptscriptstyle (2)}(\tau_f)
   \hat{U}_{\scriptscriptstyle D}(\tau_f,\tau_{\rm in}) |\Psi\rrangle,
\end{eqnarray}
with the expectation value to be taken over the initial system state
$|\psi(\tau_\mathrm{in})\rangle$, $\llangle \cdot \rrangle$ refers to the
irreducible part, and the time evolution operator reads 
\begin{equation}\label{eq:UD}
   \hat{U}_{\scriptscriptstyle D} (\tau_f,\tau_{\rm in})
   =\mathcal{T} \exp\Bigl[-\frac{i}{\hbar}\int_{\tau_{\rm in}}^{\tau_f} d\tau' 
   \hat{H}_\mathrm{sd} (\tau')\Bigr],
\end{equation}
with $\mathcal{T}$ denoting time-ordering.  Evaluating \eqref{eq:Pirrab} to
lowest relevant order in the coupling, we find
\begin{widetext}
\begin{align}
   P^\mathrm{irr}_{\alpha\beta}&= \frac{(-i)^2}{\hbar^2}
   \int_{\tau_{\rm in}}^{\tau_f} \!\!d\tau'\!
   \int_{\tau_{\rm in}}^{\tau'} \!\!d\tau''\!
   \,\llangle \Psi| [[\memproj_{\alpha}^{\scriptscriptstyle (1)}(\tau_f)
   \memproj_{\beta}^{\scriptscriptstyle (2)}(\tau_f),
   \hat{H}_\mathrm{sd}(\tau')], \hat{H}_\mathrm{sd}(\tau'')] | \Psi \rrangle
   \nonumber\\
   &= \frac{(-i)^2}{\hbar^2}
   \int_{\tau_{\rm in}}^{\tau} \!\!d\tau'
   \int_{\tau_{\rm in}}^{\tau'} \!\!d\tau''
   \bigl[
   \llangle \psi| \Oo(\tau') \Oo(\tau'') |\psi\rrangle
   \,
   \langle \phi_\mathrm{in}^{\scriptscriptstyle (1)}
   |\memproj_{\alpha}^{\scriptscriptstyle (1)}(\tau_f)
   \B^{\scriptscriptstyle (1)}(\tau'') |\phi_\mathrm{in}^{\scriptscriptstyle (1)}\rangle
   \,
   \langle \phi_\mathrm{in}^{\scriptscriptstyle (2)}
   |\memproj_{\beta}^{\scriptscriptstyle (2)}(\tau_f)
   \B^{\scriptscriptstyle (2)}(\tau') |\phi_\mathrm{in}^{\scriptscriptstyle (2)}\rangle
   \nonumber\\ \noalign{\vspace{-8pt}}
   &\hspace{97pt}-
   \llangle \psi| \Oo(\tau'') \Oo(\tau') |\psi\rrangle
   \,
   \langle \phi_\mathrm{in}^{\scriptscriptstyle (1)}|\B^{\scriptscriptstyle (1)}(\tau'')
   \memproj_{\alpha}^{\scriptscriptstyle (1)}(\tau_f) 
   |\phi_\mathrm{in}^{\scriptscriptstyle (1)}\rangle
   \,
   \langle \phi_\mathrm{in}^{\scriptscriptstyle (2)}i
   |\memproj_{\beta}^{\scriptscriptstyle (2)}(\tau_f)
   \B^{\scriptscriptstyle (2)}(\tau') |\phi_\mathrm{in}^{\scriptscriptstyle (2)}\rangle
   \nonumber\\ \noalign{\vspace{-4pt}}
   &\hspace{97pt}-
   \llangle \psi| \Oo(\tau') \Oo(\tau'') |\psi\rrangle
   \,
   \langle \phi_\mathrm{in}^{\scriptscriptstyle (1)}i
   |\memproj_{\alpha}^{\scriptscriptstyle (1)}(\tau_f)
   \B^{\scriptscriptstyle (1)}(\tau'') |\phi_\mathrm{in}^{\scriptscriptstyle (1)}\rangle
   \,
   \langle \phi_\mathrm{in}^{\scriptscriptstyle (2)}|\B^{\scriptscriptstyle (2)}(\tau')
   \memproj_{\beta}^{\scriptscriptstyle (2)}(\tau_f) i
   |\phi_\mathrm{in}^{\scriptscriptstyle (2)}\rangle
   \nonumber\\ \noalign{\vspace{-4pt}}
   &\hspace{97pt}+
   \llangle \psi| \Oo(\tau'') \Oo(\tau') |\psi\rrangle
   \,
   \langle \phi_\mathrm{in}^{\scriptscriptstyle (1)}|\B^{\scriptscriptstyle (1)}(\tau'')
   \memproj_{\alpha}^{\scriptscriptstyle (1)}(\tau_f) 
   |\phi_\mathrm{in}^{\scriptscriptstyle (1)}\rangle
   \,
   \langle \phi_\mathrm{in}^{\scriptscriptstyle (2)}|\B^{\scriptscriptstyle (2)}(\tau')
   \memproj_{\beta}^{\scriptscriptstyle (2)}(\tau_f) 
   |\phi_\mathrm{in}^{\scriptscriptstyle (2)}\rangle
   \bigr], \label{eq:Pirrexpr}
\end{align}
\end{widetext}
where we made sure that the first memory interacts with the system at the
earlier time $\tau''$. For a slow system dynamics and exploiting that
$\B^{\scriptscriptstyle (j)}(\tau)|\phi_\mathrm{in}^{\scriptscriptstyle
(j)}\rangle\neq 0$ only for $\tau \approx \tau_j$, we can replace $\Oo(\tau'')
\rightarrow \Oo(\tau_1)$ and $\Oo(\tau') \rightarrow \Oo(\tau_2)$. We make use
of the standard definitions for the symmetrized and anti-symmetrized
irreducible correlators (with $[\cdot,\cdot]$ and $\{ \cdot,\cdot\}$ denoting
the usual commutator and anti-commutator)
\begin{align} \label{eq:RSOO}
   \mathcal{R}S^\mathrm{irr}_{\Ooi\Ooi}(\tau_1,\tau_2)
   &= \bigl\llangle \{\Oo(\tau_1),
                                 \Oo(\tau_2)\}\bigr\rrangle/2,
   \\ \label{eq:ISOO}
   \mathcal{I}S^\mathrm{irr}_{\Ooi\Ooi}(\tau_1,\tau_2)
   &= -i\bigl\llangle [\Oo(\tau_1),
                                  \Oo(\tau_2)]\bigr\rrangle,
\end{align}
to arrive at the final result
\begin{align}\label{eq:Pabweak}
   P_{\alpha\beta}^\mathrm{irr} (\tau_{21})
   &=
       \mathcal{I}S^{\scriptscriptstyle (1)}_{\mathrm{det},\alpha}
       \mathcal{I}S^{\scriptscriptstyle (2)}_{\mathrm{det},\beta}
       \mathcal{R}S^\mathrm{irr}_{\Ooi\Ooi}(\tau_1,\tau_2) \\
     \nonumber
   &\quad+
       \mathcal{R}S^{\scriptscriptstyle (1)}_{\mathrm{det},\alpha}
       \mathcal{I}S^{\scriptscriptstyle (2)}_{\mathrm{det},\beta}
       \mathcal{I}S^\mathrm{irr}_{\Ooi\Ooi}(\tau_1,\tau_2),
\end{align}
with the detector response functions 
\begin{align}
   \mathcal{R}S_{\mathrm{det},\alpha}^{\scriptscriptstyle (j)}
   &= -\frac{1}{2\hbar}\int_{\tau_{\rm in}}^{\tau_f}d\tau
   \langle \phi^{\scriptscriptstyle (j)}_\mathrm{in}|
        \{\memproj_{\alpha}^{\scriptscriptstyle (j)}(\tau_f), \B(\tau)\}|
           \phi^{\scriptscriptstyle (j)}_\mathrm{in} \rangle,
   \label{eq:RSdet}\\
   \mathcal{I}S_{\mathrm{det},\alpha}^{\scriptscriptstyle (j)}
   &=\frac{-i}{\hbar} \int_{\tau_{\rm in}}^{\tau_f}d\tau
   \langle \phi^{\scriptscriptstyle (j)}_\mathrm{in}|
        [\memproj_{\alpha}^{\scriptscriptstyle (j)}(\tau_f), \B(\tau)]|
           \phi^{\scriptscriptstyle (j)}_\mathrm{in} \rangle.
   \label{eq:ISdet}
\end{align}
The symbols $\mathcal{R}$ and $\mathcal{I}$ address symmetrized and
anti-symmetrized quantities (or equivalently, up to factors of 2, real- and
imaginary parts).

In a situation where the full information $P_{\alpha\beta}$ can be extracted
from the memories, the individual correlators $\mathcal{R}S^{\rm
irr}_{\Ooi\Ooi}$ and $\mathcal{I}S^{\rm irr}_{\Ooi\Ooi}$ can be obtained from
(\ref{eq:Pabweak}) by combining two different probabilities, e.g., using
$P^{\rm irr}_{\alpha\beta}$ and $P^{\rm irr}_{\bar\alpha\beta}$ one obtains
\begin{align}\label{eq:RSOOf}
   \mathcal{R}S^{\rm irr}_{\Ooi\Ooi}
   &=
   [\mathcal{R}S^{\scriptscriptstyle (1)}_{{\rm det},\bar\alpha}
                                         P^{\rm irr}_{\alpha\beta}
   - \mathcal{R}S^{\scriptscriptstyle (1)}_{{\rm det},\alpha}
                                         P^{\rm irr}_{\bar\alpha\beta} ]/
   D^{\scriptscriptstyle (1)}\,  
   \mathcal{I}S^{\scriptscriptstyle (2)}_{{\rm det},\beta},
   \\ \label{eq:ISOOf}
   \mathcal{I}S^{\rm irr}_{\Ooi\Ooi}
   &= [-\mathcal{I}S^{\scriptscriptstyle (1)}_{{\rm det},\bar\alpha}
                                         P^{\rm irr}_{\alpha\beta}
   + \mathcal{I}S^{\scriptscriptstyle (1)}_{{\rm det},\alpha}
                                         P^{\rm irr}_{\bar\alpha\beta} ]/
   D^{\scriptscriptstyle (1)}\, 
   \mathcal{I}S^{\scriptscriptstyle (2)}_{{\rm det},\beta},
\end{align}
with $D^{\scriptscriptstyle (1)} =\mathcal{R}S^{\scriptscriptstyle (1)}_{{\rm
det},\bar\alpha} \mathcal{I}S^{\scriptscriptstyle (1)}_{{\rm det}, \alpha}
-\mathcal{R}S^{\scriptscriptstyle (1)}_{{\rm det},\alpha}
\mathcal{I}S^{\scriptscriptstyle (1)}_{{\rm det},\bar\alpha}$. Alternatively,
one may have preferential access to combinations of probabilities
$P_{\alpha\beta}$ (see Sec.\ \ref{sec:charge_exp} for an example) or make use
of specific detector properties, see below and the appendix for examples.

A generic choice for the quantum memories are qubit devices that couple to the
system via the Hamiltonian $\hat{H}_\mathrm{sd}(\tau) = \Omega(\tau)
\hat{\sigma}_x \Oo$, where the coupling $\Omega(\tau)$ is switched on during a
short time $\delta \tau$ around $\tau_j$. Assuming initial states
\begin{align}\label{eq:q_in}
   |\phi^{\scriptscriptstyle (j)}_\mathrm{in}\rangle = [|\varphi_0\rangle 
   + e^{i\theta^{\scriptscriptstyle (j)}}|\varphi_1\rangle]/\sqrt{2},
\end{align}
a system residing in a state $|n\rangle$ with eigenvalue $O_n$ will rotate the
qubit around the $x$-axis by $\delta \vartheta\, O_n$ with
$\delta\vartheta=\Omega\delta \tau/\hbar$. The response functions
(\ref{eq:RSdet}) and (\ref{eq:ISdet}) involve integrals of the type
\begin{align}
   \int_{\tau_{\rm  in}}^{\tau_{f}} d\tau
   \langle \phi_\mathrm{in}^{\scriptscriptstyle (j)}| \memproj_\alpha(\tau_f)
   \Omega(\tau)\hat{\sigma}_x| \phi_\mathrm{in}^{\scriptscriptstyle (j)} \rangle =
   \delta\vartheta\, e^{i(1-2\alpha)\theta^{\scriptscriptstyle (j)}}
\end{align}
and we obtain the response functions
\begin{align}
   \label{eq:q_res_r}
   \mathcal{R}S_\mathrm{det,0}^{\scriptscriptstyle (j)} 
   &=\delta\vartheta\,\cos\theta^{\scriptscriptstyle (j)},
   \quad
   \mathcal{R}S_\mathrm{det,1}^{\scriptscriptstyle (j)} 
   =\delta\vartheta\,\cos\theta^{\scriptscriptstyle (j)},\\
   \label{eq:q_res_i}
   \mathcal{I}S_\mathrm{det,0}^{\scriptscriptstyle (j)} 
   &=2\delta\vartheta\,\sin\theta^{\scriptscriptstyle (j)},
   \quad
   \mathcal{I}S_\mathrm{det,1}^{\scriptscriptstyle (j)} 
   =-2\delta\vartheta\,\sin\theta^{\scriptscriptstyle (j)}.
\end{align}
Choosing $\theta^{\scriptscriptstyle (2)} = \pi/2$, i.e., polarizing the
second memory along the $y$-axis, we can directly find the correlator
$\mathcal{R} S^\mathrm{irr}_{\Ooi\Ooi}$ ($\mathcal{I}
S^\mathrm{irr}_{\Ooi\Ooi}$) from the memory correlator
$P^\mathrm{irr}_{\alpha\beta}$ by choosing $\theta^{\scriptscriptstyle (1)} =
\pi/2$ ($\theta^{\scriptscriptstyle (1)} = 0$). Alternatively, we may use the
results (\ref{eq:RSOOf}) and (\ref{eq:ISOOf}) and $\theta^{\scriptscriptstyle
(1)} = \theta^{\scriptscriptstyle (2)} = \pi/4$ to find (we choose $\alpha =
0$, $\bar{\alpha} = 1$, and $\beta = 0$)
\begin{align}\label{eq:RS_q}
   \mathcal{R} S^\mathrm{irr}_{\Ooi\Ooi} = \frac{P^\mathrm{irr}_{00} -
   P^\mathrm{irr}_{10}}{4 \delta\vartheta^2}, \\ \label{eq:IS_q} \mathcal{I}
   S^\mathrm{irr}_{\Ooi\Ooi} = -\frac{P^\mathrm{irr}_{00} +
   P^\mathrm{irr}_{10}}{2 \delta\vartheta^2}.
\end{align}
Note that the sums $P^\mathrm{irr}_{00} + P^\mathrm{irr}_{10}$ and
$P^\mathrm{irr}_{00} + P^\mathrm{irr}_{01}$ contain very different types of
information, once a correlation, the other time only a mean value, as
discussed in more detail in Sec.\ \ref{sec:charge_exp}.  Hence, we find that
the delayed measurement of the two quantum memories provides the symmetrized
and anti-symmetrized correlators (\ref{eq:RSOO}) and (\ref{eq:ISOO}).
Finally, we note that a weak linear coupling between the system observable and
the detector/memory variable canonically conjugated to the detector readout
provides a more effective entanglement. Such a von Neumann like interaction
allows to produce a strong entanglement and a strong measurement even for weak
coupling if sufficient time is available for the entanglement
process\cite{gurvitz:06}.

\subsection{Strong measurement}\label{sec:gen_sc}

We are now going to show that a {\it strong system--detector coupling}
naturally leads to projective correlators. Consider a situation where the
system operator $\Oo$ is measured via entanglement with two quantum
memories.  We first consider an operator $\Oo$ with a non-degenerate
spectrum and comment on the general case in the end.  A strong coupling
between the system and the memory implies that the memory states
$|\phi^{\scriptscriptstyle (j)}_{n}\rangle$ after interaction with the system
in state $|n\rangle$ are fully distiguishable, i.e, 
\begin{align}
   \langle \phi^{\scriptscriptstyle (j)}_{m} 
          |\phi^{\scriptscriptstyle (j)}_{n}\rangle =\delta_{nm}.
\end{align}
The observables $\A^{\scriptscriptstyle (j)}$ distinguish between
memory eigenstates $|\varphi_{\alpha}^{\scriptscriptstyle (j)} \rangle$ and we
assume a one-to-one relation with the evoluted memory states
$|\phi^{\scriptscriptstyle (j)}_{n} \rangle$,
\begin{align} \label{eq:kroneckerscattering}
  |\phi^{\scriptscriptstyle (j)}_{n}\rangle=
  |\varphi_{\alpha_n}^{\scriptscriptstyle (j)} \rangle,
\end{align}
with $\alpha_n \neq \alpha_m$ for $n \neq m$ (otherwise, the observable
$\A^{\scriptscriptstyle (j)}$ measures linear combinations of 
eigenstates of $\Oo$ and thus is not suitable for a measurement of this
observable). Under these (strong coupling) conditions, the amplitudes
$s_n^{\alpha}$ reduce to $s_n^{\alpha} = \delta_{\alpha \alpha_n}$ (for $j=1$,
$s_m^{\beta} = \delta_{\beta \beta_m}$ for the second memory).  In order to
describe the strong coupling situation it is favorable to proceed with the
expression (\ref{eq:Pab}) and we obtain the result
\begin{align} \label{eq:Panbm}
   P_{\alpha_n \beta_m}(\tau_{21})= \bigl| U_{mn}(\tau_{21})\psi_n(\tau_1) 
   \bigr|^2.
\end{align}
The right hand side of the above expression is nothing but the projected
system correlator
\begin{align}\label{eq:SPP}
   S^{P}_{\sysproji_{n}\sysproji_{m}}(\tau_{21})
   &=\sum_l \langle l | \sysproj_m(\tau_2)\sysproj_n(\tau_1)
   \rho_0 \sysproj_n (\tau_1) |l\rangle\\ \nonumber
   &= |U_{mn}(\tau_{21})\psi_n(\tau_1)|^2,
\end{align}
with the projected density matrix $\rho^{P}(\tau_1) =
\sum_{k}\sysproj_{k}(\tau_1)$ $\rho_0 \sysproj_{k}(\tau_1)$ and the projectors
$\sysproj_{k}=|k\rangle \langle k|$ onto different system states, see Eq.\
(\ref{eq:Psi}).  The  projected correlators $S^{P}_{\sysproji_{n}
\sysproji_{m}} (\tau_{21})$ are easily combined into the desired two-time
correlator $S^{P}_{\Ooi\Ooi}(\tau_{21})$
\begin{align}\label{eq:SPOO}
   S_{\Ooi\Ooi}^P(\tau_{21}) &= {\rm Tr} [\Oo 
   (\tau_2) \Oo (\tau_1) \rho^P (\tau_1)] \\ \nonumber
   &= \sum_{nm} O_n O_m S^{P}_{\sysproji_{n}\sysproji_{m}}(\tau_{21})\\  \nonumber
   &= \sum_{nm} O_n O_m P_{\alpha_n \beta_m}(\tau_{21}),
\end{align}
thereby establishing the general result relating the system correlator to the
memory readings. A further simplification can be achieved if the eigenvalues
of $\A^{\scriptscriptstyle (j)}$ and $\Oo$ obey a linear relation $O_n =
\eta^{\scriptscriptstyle (j)} a^{\scriptscriptstyle (j)}_{\alpha_n}$, in which
case we obtain the simple result
\begin{align} \nonumber
    S^{P}_{\Ooi\Ooi}(\tau_{21})
    & = \sum_{nm} 
        \eta^{\scriptscriptstyle (1)} a^{\scriptscriptstyle (1)}_{\alpha_n}
        \eta^{\scriptscriptstyle (2)} a^{\scriptscriptstyle (2)}_{\beta_m}
               P_{\alpha_n \beta_m}(\tau_{21}) \\
   & = \eta^{\scriptscriptstyle (1)}\eta^{\scriptscriptstyle (2)}
   \langle \A^{\scriptscriptstyle (1)}
   \A^{\scriptscriptstyle (2)}\rangle,
   \label{eq:AASPOO}
\end{align}
directly relating the projected system correlator to the memory correlator
$\langle \A^{\scriptscriptstyle (1)} \A^{\scriptscriptstyle (2)}\rangle$.

These results can easily be generalized to the case where the observable $\Oo$
involves a degenerate spectrum: The system evolution of the memories is
conditioned on the eigenvalue $O_n$ rather than the state $|n\rangle$ of the
system. Degenerate eigenstates produce equal evolutions of the memories,
implying that $P_{\alpha_n \beta_m}(\tau_{21})$ has to be replaced by
$P_{\alpha_O \beta_{O'}}(\tau_{21}) = \sum_{\{m|O_m=O'\}}| \sum_{\{n|O_n=O\}}
U_{mn}(\tau_{21})\psi_n(\tau_1)|^2$ and $\rho^P(\tau_1)$ is substituted by
$\sum_O \sysproj_O(\tau_1) \rho_0 \sysproj_O(\tau_1)$ with the projection
operator $\sysproj_O = \sum_{\{n|O_n = O\}}|n\rangle \langle n|$. With these
replacements, the above results remain valid also for a degenerate spectrum.

Hence, the perfect entanglement between the system and the memories arising
due to strong coupling is equivalent to a von Neumann projection applied to
the system. While no back action is apparent on the level of the system
dynamics, the strong back action of this maximal entanglement with the
memories manifests itself in a strong change of the system’s density matrix
when tracing over the memories.  

The realization of a strong measurement as described above requires equal or
more memory states $|\varphi_\alpha\rangle$ than system states $|n\rangle$,
hence quantum memories with dimensionality $d$ or qudits are required. Besides
the obvious difficulty in realizing qudit memories, their coupling to the
system in order to serve as a measurement device is equally nontrivial. As an
alternative, we discuss a measurement scheme involving qubit registers,
instead.

Such an alternative setup implementing a strong measurement involves a weak
system--detector coupling but invokes multiple measurements. We replace the
strongly coupled qudit memories by weakly coupled qubit registers with $J$
qubits each, probing the system state close to $\tau_1$ and $\tau_2$ within a
short time interval $\delta \tau$. For a strong measurement, $J$ is chosen
sufficiently large to distinguish the different eigenvalues $O_n$ of the
system. Assuming the system not to change during the interaction time $J
\delta \tau$ with one register, the final state of system and memory registers
is of the same form as in Eq.\ \eqref{eq:Psif} with the outgoing individual
memory states $|\phi^{\scriptscriptstyle (1)}_n\rangle$ and
$|\phi^{\scriptscriptstyle (2)}_m\rangle$ replaced by the outgoing register
states $|\Phi^{\scriptscriptstyle (1)}_n\rangle=\prod_{j=1}^J
|\phi_{n}^{\scriptscriptstyle (j)}\rangle$ and $|\Phi^{\scriptscriptstyle
(2)}_m\rangle = \prod_{j=J+1}^{2J} |\phi_{m}^{\scriptscriptstyle (j)}\rangle$.
Making use of Born's rule, we obtain the probabilities
$P_{\alpha\beta}^{\mu\nu} (\tau_{21})$ for finding $\mu$ ($\nu$) qubits of the
first (second) register in states $\alpha \in \{0,1\}$ ($\beta \in \{0,1\}$),
\begin{align}\nonumber
   P_{\alpha\beta}^{\mu\nu}
   \!=
   {\binom{J}{\mu} \binom{J}{\nu}}
   \sum_m \Bigl| \sum_n & (s_m^\beta)^\nu (s_m^{\bar\beta})^{J-\nu} U_{mn}
    (s_n^\alpha)^\mu (s_n^{\bar\alpha})^{J-\mu} \psi_n
          \Bigr|^2\!\!,
\end{align}
with $s_n^{\bar \alpha} = s_n^{1-\alpha}$ and $s_n^{\bar \beta} =
s_n^{1-\beta}$.  Introducing the conditional probability $P_\alpha^\mu(n) =
\binom{J}{\mu} |s_n^\alpha|^{2\mu} |s_n^{\bar\alpha}|^{2(J-\mu)}$ for
measuring $\mu$ of the $J$ qubits in the state $\alpha$ after interaction with
the system in state $|n\rangle$, we can separate the system- and detector
response in the above equation,
\begin{align}\label{eq:PabmnRP}
   P_{\alpha\beta}^{\mu\nu}
   =
   \sum_{n,n',m} [U_{mn} \psi_n U_{mn'}^\ast \psi_{n'}^\ast]
           P_\alpha^\mu(n,n') P_{\beta}^\nu(m)
\end{align}
with the `off-diagonal' conditional probabilities $P_\alpha^\mu(n,n') =
\binom{J}{\mu} (s_n^\alpha {s_{n'}^\alpha}^\ast)^{\mu} (s_n^{\bar\alpha}
{s_{n'}^{\bar\alpha}}^\ast)^{(J-\mu)}$ (note that $P_\alpha^\mu(n,n) =
P_\alpha^\mu(n)$).  The conditional probabilities $P_{\alpha}^\mu(n)$ depend
only on the eigenvalue $O_n$ of the state $|n\rangle$ (and not on the state
$|n\rangle$ itself).  We then have to distinguish two cases: (i) all system
eigenvalues are non-degenerate, i.e., $O_n \neq O_{n'}$ for $n\neq n'$,
(ii) there are degenerate system states.

In the non-degenerate case (i) and for a strong measurement, the probability
distributions $P_{\alpha}^\mu(n)$ and  $P_{\alpha}^\mu(n')$ for different $n
\neq n'$ do not overlap as functions of $\mu$ (this is the very definition of
this measurement being a strong one) and the `off-diagonal' elements
$P_\alpha^\mu(n,n')$ for $n'\neq n$ are suppressed, as follows from the
relation $|P_\alpha^\mu(n,n')| = \sqrt{P_{\alpha}^\mu(n) P_{\alpha}^\mu(n')}$.
We then can simplify the expression (\ref{eq:PabmnRP}) to the form
$P_{\alpha\beta}^{\mu\nu} \approx \sum_{n,m} |U_{mn} \psi_n|^2 P_\alpha^\mu(n)
P_{\beta}^\nu(m)$. The register correlators (replacing the distribution
functions $P_{\alpha\beta}$) take the form
\begin{align}\label{eq:Sab}
   S_{\alpha\beta}(\tau_{21}) = \langle \Psi_f| \sum_{j=1}^J
   \hat{p}_\alpha^{\scriptscriptstyle (j)} \sum_{j=J+1}^{2J}
   \hat{p}_\beta^{\scriptscriptstyle (j)}|\Psi_f\rangle =
   \sum_{\mu,\nu} \mu \nu P_{\alpha\beta}^{\mu\nu}\\
   \nonumber
   = \sum_{n,m} |U_{mn} \psi_n|^2  
   \sum_{\mu} \mu P_\alpha^\mu(n)
   \sum_{\nu} \nu P_\beta^\nu(m).
\end{align}
Assuming again a linear system--qubit coupling $\hat{H}_\mathrm{sd}(\tau) =
\Omega(\tau)\hat{\sigma}_x \Oo$ that rotates the qubits by an angle
$\delta\vartheta\, O_n$ around the $x$-axis, the evolution
\begin{align}\label{eq:u}
  \hat{u}_n = \begin{pmatrix}
   \cos(\delta\vartheta\, O_n) & - i \sin(\delta\vartheta\, O_n)\\
   i\sin(\delta\vartheta\, O_n) & \cos(\delta\vartheta\, O_n) \end{pmatrix}
\end{align}
produces the memory states $|\phi^{\scriptscriptstyle (j)}_n\rangle = \hat{u}_n |\phi^{\scriptscriptstyle
(j)}_\mathrm{in}\rangle$, where we again assume initial states polarized in
the $xy$-plane, see (\ref{eq:q_in}). The probabilities $P_{\alpha}(n)$ for an
individual qubit to reside in state $\alpha = 0,~1$ are given by
\begin{align}
   P_0(n) = 1-P_1(n)&
   \approx \frac{1}{2} + (1-2\alpha) \delta\vartheta\, O_n \sin\theta.
   \label{eq:prob_q}
\end{align}
With the qubits initially polarized along the $y$-axis, i.e., $\theta^{\scriptscriptstyle
(j)} = \pi/2$, we define the register's `magnetizations'
\begin{align}\label{eq:Mn}
   M(n) = \sum_\mu  \mu [P_0^\mu(n) - P_1^\mu(n)] =2 J \delta\vartheta\, O_n,
\end{align}
where we have made use of (\ref{eq:prob_q}) in the last equation.  The
combination $S_{11}-S_{10}-S_{01}+S_{00}$ then involves the product of
register polarizations $M(n) M(m)$ and using the relation
$S_{\Ooi\Ooi}^P(\tau_{21}) = \sum_{nm} O_n O_m |U_{mn}\psi_n|^2$, see Eq.\
(\ref{eq:SPOO}), we can relate the system correlator to the register
correlators via
\begin{align}\label{eq:SPOOreg}
   S_{\Ooi\Ooi}^P(\tau_{21}) = \frac{S_{11}-S_{10}-S_{01}+S_{00}}
                                      {4 J^2 \delta\vartheta^2}.
\end{align}
Note the normalization $S_{11}+S_{10}+S_{01}+S_{00} = J^2$, which follows from
replacing $M(n) = \sum_\mu  \mu [P_0^\mu(n) - P_1^\mu(n)]$ by $\Sigma(n) =
\sum_\mu  \mu [P_0^\mu(n) + P_1^\mu(n)] =\langle \mu \rangle_0 + \langle \mu
\rangle_1 = J$.

In the degenerate case (ii), the distribution functions separate only for
states with different eigenvalues $O_n\neq O_{n'}$, implying that
$P_\alpha^\mu(n,n') \sim 0$, while for degenerate eigenvalues with $O_n =
O_{n'}$, $P_\alpha^\mu(n,n') = P_{\alpha}^\mu(n)$. The probabilities
$P_{\alpha\beta}^{\mu\nu}$ then are given by
\begin{align}\nonumber
   P_{\alpha\beta}^{\mu\nu}
   \approx \sum_{m,n}  \sum_{\{n'|O_{n'}=O_n\}} 
           [U_{mn} \psi_n U_{mn'}^\ast \psi_{n'}^\ast]
           P_\alpha^\mu(n) P_{\beta}^\nu(m).
\end{align}
On the other hand, the projected density matrix $\rho^P(\tau_1)$ appearing in
the system correlator $S_{\Ooi\Ooi}^P(\tau_{21})$ involves the projectors
$\hat{P}_O = \sum_{\{n|O_n = O\}}|n\rangle \langle n|$, such that the result
(\ref{eq:SPOOreg}) remains unchanged.

Summarizing, we have seen that the measurement scheme invoking entanglement
with quantum memories and their delayed measurement provides us, in the weak-
and strong measurement limits, with the same results as obtained via the
traditional route using an intermediate von Neumann projection.  Furthermore,
these results have been obtained within a unified description starting from
the same initial formula in the form (\ref{eq:Pabdef}) or (\ref{eq:Pab}).

The above general theoretical considerations are rather non-trivial to
implement in a practical situation, as the preparation, entanglement, and
measurement of many quantum memories is often a non-trivial task. The
implementation of these ideas is less demanding, though still challenging,
when considering specific examples. Indeed, quantum memories for delayed
measurement are naturally provided in a scattering geometry, where individual
scattered particles take the role of flying qubits.  In the following, we
focus on a specific example in mesoscopic physics, the measurement of a charge
correlator of a quantum dot by scattering electrons in a nearby quantum point
contact. We first analyze the situation for a simple two-state system with
charges $\eigQ = ne$, $n=0,1$, and then extend these considerations to
arbitrary charge states.

\section{Charge correlator measurement}\label{sec:charge}

\begin{figure}[h]
\begin{center}
\includegraphics[width=7cm]{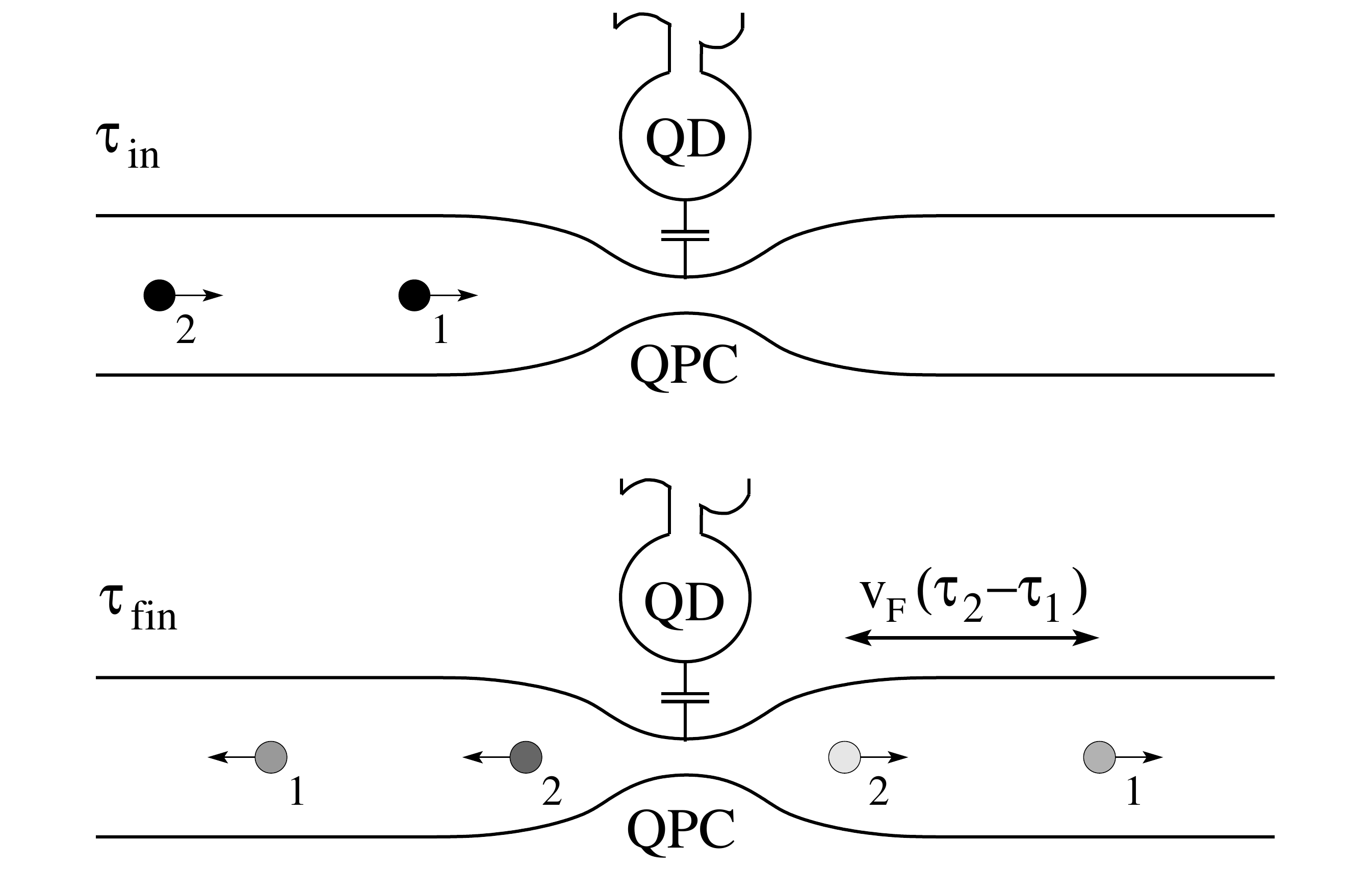}
\end{center}
\caption{\label{fig:meso_setup} Quantum dot system (QD) measured by a
capacitively coupled quantum point contact (QPC): Single-electron pulses
incident on the QPC from the left (at $\tau_\mathrm{in}$) are either
transmitted (with amplitude $t$) or reflected (amplitude $r$); the outgoing
Lippmann-Schwinger wave functions describe flying qubits without own dynamics
and serve as quantum memories. Two pulses separated in time by $\tau_2 - \tau_1$ are
needed to measure the two-time correlator of the dot's charge.  After
scattering at the QPC, the two electrons (flying qubits) are entangled with
the quantum dot system and carry information on its dynamics.  Simultaneous
detection of the two scattered electrons (at $\tau_\mathrm{fin}$), e.g., a
distance $v_{\rm\scriptscriptstyle F} (\tau_2-\tau_1)$ away with both positions on
the right of the QPC, provides information on the two-time charge correlator.
}
\end{figure}

We consider a classic problem \cite{classic_qpc_qd}, the charge $\Q$ dynamics
of a quantum dot (QD) (attached to leads or coupled to another dot in an
isolated double-dot system) measured by a nearby quantum point contact (QPC),
see Fig.\ \ref{fig:meso_setup} and Ref.\ \onlinecite{oehri:14} for a recent
discussion of this system. Here, we want to characterize the dot's dynamics by
its two-time charge correlator. In this system, the measurement is executed by
the electrons which are transmitted across/reflected by the QPC with
probabilities that depend on the dot's charge state.  For the present
discussion, it is convenient to view the QPC current as a sequence of
individual electron pulses; during recent years, this theoretical idea
\cite{levitov:96,lebedev:05,keel:06} has progressed to an experimental reality
\cite{feve:07,fletcher:13,bocquillon:13}, opening the new field of electron
quantum optics \cite{eqo}.  The quantum memories then can be viewed as flying
qubits, individual electron pulses arriving at the QPC at times $\tau_1$ and
$\tau_2$ that probe the charge state of the QD through the capacitive coupling
between the QD and the QPC, see Fig.\ \ref{fig:meso_setup}.

The memory states $|\varphi_\alpha^{\scriptscriptstyle (j)}\rangle$ are the
two scattered states where the electron is reflected ($\alpha = \textrm{r}$)
or transmitted ($\alpha = \textrm{t}$), i.e., the outgoing state is given by
$|\phi_{n}^{\scriptscriptstyle (j)} \rangle = t_n
|\varphi_\mathrm{t}^{\scriptscriptstyle (j)} \rangle + r_n
|\varphi_\mathrm{r}^{\scriptscriptstyle (j)} \rangle$ with scattering
coefficients $t_n \leftrightarrow s_n^\mathrm{t}$ and $r_n \leftrightarrow
s_n^\mathrm{r}$ depending on the charge state of the system.  We assume well
separated single-electron pulses and an evolution of the scattered waves
$|\varphi_\mathrm{r,t}^{\scriptscriptstyle (j)}\rangle$ emanating from the QPC
at times $\tau_j$ that preserves the corresponding system information, in
particular, $\langle \phi_{m}^{\scriptscriptstyle (2)} |
\phi_{n}^{\scriptscriptstyle (1)} \rangle = 0$. This allows us to envision an
individual detection of the electrons \cite{bocquillon:13,thalineau:14}.  In the
final readout, the flying qubits are detected on the right or left side of the
QPC, telling whether the two electrons have been transmitted (with probability
$P_\mathrm{tt}$), reflected ($P_\mathrm{rr}$) or mixed ($P_\mathrm{tr}$ and
$P_\mathrm{rt}$). These probabilities then contain the information about the
two-time charge correlator of the QD.

Formally, such a final-state analysis corresponds to measuring the (charge)
operators $\A^{\scriptscriptstyle (j)} = \sum_\alpha
a_\alpha^{\scriptscriptstyle (j)} \memproj_\alpha^{\scriptscriptstyle (j)}$,
with the projectors $\memproj_\alpha^{\scriptscriptstyle (j)} =
|\varphi_\alpha^{\scriptscriptstyle (j)}\rangle \langle
\varphi_\alpha^{\scriptscriptstyle (j)}|$ providing the transmitted ($\alpha =
\mathrm{t}$) or reflected ($\alpha = \mathrm{r}$) components of the $j$-th
electron. The eigenvalues $a_\alpha^{\scriptscriptstyle (j)}$ depend on the
measured charge, e.g., $a_\mathrm{t}^{\scriptscriptstyle (j)} = 1$ and
$a_\mathrm{r}^{\scriptscriptstyle (j)} = 0$ if the transmitted charge is
measured on the right of the QPC, while $a_\mathrm{r}^{\scriptscriptstyle (j)}
= 1$ and $a_\mathrm{t}^{\scriptscriptstyle (j)} = 0$ if the reflected charge
is measured on the left. Both types of measurements provide us with the same
probability distributions (\ref{eq:Pabdef}) or (\ref{eq:Pab}) with
$\alpha,~\beta = \mathrm{r,~t}$. 

In the following, we first analyze the case of a QD with binary charge states
$|0\rangle$ and $|1\rangle$ and eigenvalues $Q_0 = 0$, $Q_1 = 1$, $\Q =
|1\rangle \langle 1|$ (defined in units of $e$) and an initial state
$|\psi(\tau_\mathrm{in})\rangle = \psi_0(\tau_\mathrm{in}) |0\rangle +
\psi_1(\tau_\mathrm{in}) |1\rangle$, see Eq.\ (\ref{eq:Psi}). Second, we
generalize the discussion to a QD with multiple charge states as described by
the charge operator $\hat{Q} = \sum_n Q_n |n\rangle \langle n|$. While
two-state quantum memories (qubits) are always sufficient for a complete
description when the measurement is weak, for a strong measurement, a QD with
multiple charge states will require the use of qubit registers, i.e., finite
trains of electron pulses.

\subsection{Weak measurement}\label{sec:charge_wc}

For the case of a weak measurement, we can make immediate use of the general
results Eqs.\ (\ref{eq:Pabweak}), (\ref{eq:RSOOf}), and (\ref{eq:ISOOf}) by
replacing $\Oo \to \Q$ and choosing the values $\alpha,~\beta$ equal to
$\mathrm{r}$ and $\mathrm{t}$. It remains to determine the detector response
functions (\ref{eq:ISdet}) and (\ref{eq:RSdet}). We consider a linear
system--detector coupling of the form $H_\mathrm{sd} = e^2 \q\Q/C = \vv \Q$
with $\q$ the charge on the QPC. Furthermore, we parametrize the scattering
matrices for the system in states $|0\rangle$ and $|1\rangle$ by $t_0 =
\sqrt{T} \, e^{i\theta}$, $r_0 = \sqrt{R} \, e^{i\chi}$, $t_1 = \sqrt{T -
\delta T} \, e^{i(\theta + \delta\theta)}$, $r_1 = \sqrt{R + \delta T} \,
e^{i(\chi + \delta\chi)}$, with small corrections $\delta T$, $\delta \theta$,
and $\delta \chi$. 

In order to find the detector response functions (\ref{eq:RSdet}) and
(\ref{eq:ISdet}) we replace $\B$ by $\hat{v}$ and determine the integral
$(-i/\hbar)\int_{\tau_{\rm in}}^{\tau_{f}} d\tau \langle \phi_\mathrm{in}|
\memproj_\alpha(j) (\tau_f) \hat{v}(\tau)| \phi_\mathrm{in} \rangle$ (we drop
the memory index $^{\scriptscriptstyle(j)}$ as the electrons are scattered by
the same QPC). To lowest order in $\hat v$, this can be written in the form
$\langle \phi_\mathrm{in}| \hat{u}_0^\dagger \memproj_{\alpha}
\hat{u}_1|\phi_\mathrm{in}\rangle-\langle\phi_\mathrm{in}| \hat{u}_0^\dagger
\memproj_{\alpha}\hat{u}_0|\phi_\mathrm{in}\rangle$ with $\hat{u}_0 =
e^{-i\hat{h}_0(\tau_f-\tau_{\rm in})/\hbar}$ and $\hat{u}_1 = e^{-i(\hat{h}_0 +
\hat{v})(\tau_f-\tau_{\rm in})/\hbar}$ describing the dynamics of the detector
in the Heisenberg representation in the absence and presence of a charge on
the dot, respectively.  Furthermore, we have used that $\hat{u}_1 = \hat{u}_0
\hat{u}_{\scriptscriptstyle D}$ with
\begin{align}
   \hat{u}_{\scriptscriptstyle D} = \mathcal{T} \exp\bigl(-i\int_{\tau_{\rm
   in}}^{\tau_{f}} d\tau \hat{v}(\tau)/\hbar\bigr)
\end{align}
and $\hat{v}(\tau)$ in the interaction representation. The initial memory
states $|\phi_\mathrm{in}\rangle$ evolve under $\hat{u}_1$ and $\hat{u}_0$ according to
$\hat{u}_n |\phi_\mathrm{in}\rangle = |\phi_n\rangle$ and hence $\langle
\phi_\mathrm{in}| \hat{u}_m^\dagger \memproj_{\alpha} \hat{u}_n|\phi_\mathrm{in} \rangle =
\langle\phi_m| \memproj_{\alpha} |\phi_n\rangle = s^{\alpha\ast}_m
s^\alpha_n$, where we have used that $\memproj_{\alpha} = |\varphi_\alpha
\rangle \langle \varphi_\alpha|$.  Expanding $s^{\alpha}_1$ in $\delta T$,
$\delta\theta$, and $\delta\chi$, we find that
\begin{align}\label{eq:dett}
   \frac{-i}{\hbar}\int_{\tau_{\rm  in}}^{\tau_{f}} d\tau
   \langle \phi_\mathrm{in}| \memproj_\mathrm{t}(\tau_f)
   \hat{v}(\tau)| \phi_\mathrm{in} \rangle
   &=-\frac{1}{2}\delta T+i T\delta \theta,\\
   \label{eq:detr}
   \frac{-i}{\hbar}\int_{\tau_{\rm  in}}^{\tau_{f}} d\tau
   \langle \phi_\mathrm{in}| \memproj_\mathrm{r}(\tau_f)
   \hat{v}(\tau)| \phi_\mathrm{in} \rangle
   &=\frac{1}{2}\delta T+i R\delta \chi,
\end{align}
and taking real and imaginary parts, we arrive at the final results
\begin{align}\label{eq:det_R}
   \mathcal{R}S_\mathrm{det,t} &=T \delta \theta,
   \quad
   \mathcal{R}S_\mathrm{det,r} =R \delta \chi,\\
   \label{eq:det_I}
   \mathcal{I}S_\mathrm{det,t} &=-\delta T,
   \quad
   \mathcal{I}S_\mathrm{det,r} =\delta T.
\end{align}
Using these detector response functions in the general expressions
(\ref{eq:RSOOf}) and (\ref{eq:ISOOf}) one easily arrives at the
system correlators (we choose $\alpha = \mathrm{t}$,
$\bar{\alpha} = \mathrm{r}$, and $\beta = \mathrm{t}$),
\begin{align}\label{eq:RS_c}
   \mathcal{R} S^\mathrm{irr}_{QQ} &= \frac{R(\delta\chi/\delta T)\, 
   P^\mathrm{irr}_\mathrm{tt} - T(\delta\theta/\delta T)\, P^\mathrm{irr}_\mathrm{rt}}
   {\delta T \,(R\delta\chi +T \delta\theta)}, \\ 
   \label{eq:IS_c} 
   \mathcal{I} S^\mathrm{irr}_{QQ} &= -\frac{P^\mathrm{irr}_\mathrm{tt} +
   P^\mathrm{irr}_\mathrm{rt}}
   {\delta T \,(R\delta\chi +T \delta\theta)}.
\end{align}

Alternatively, the QPC can be tuned to deliver the individual system
correlators $\mathcal{R}S^{\rm irr}_{\Qi\Qi}$ or $\mathcal{I}S^{\rm
irr}_{\Qi\Qi}$. Indeed, using a detector with high transmission, e.g., a QPC
with energetic ($E$) single-electron pulses $E \gg V_0$, $V_0$ the QPC
barrier, one easily finds that (although $\delta T \ll |\delta\theta|,
|\delta\chi|$, we have $\delta T \gg R |\delta\chi|$)
\begin{align}
   |\mathcal{R}S_\mathrm{det,t}|\gg |\mathcal{I}S_\mathrm{det,r/t}|
   \gg |\mathcal{R}S_\mathrm{det,r}|,
\end{align}
and therefore $P^\mathrm{irr}_{\mathrm{t}\beta}$ measures $\mathcal{I}S^{\rm
irr}_{\Qi\Qi}$, while $P^\mathrm{irr}_{\mathrm{r}\beta}$ measures $\mathcal{R}
S^{\rm irr}_{\Qi\Qi}$. When the detector predominantly reflects particles, e.g.,
for low energy single electron pulses with $E\ll V_0$, the situation is
reverse, $|\mathcal{R}S_\mathrm{det,r}|\gg |\mathcal{I}S_\mathrm{det,r/t}| \gg
|\mathcal{R}S_\mathrm{det,t}|$, and $\mathcal{I}S^{\rm irr}_{\Qi\Qi}$
($\mathcal{R}S^{\rm irr}_{\Qi\Qi}$) is measured by $P^\mathrm{irr}_{\mathrm{r}
\beta}$ ($P^\mathrm{irr}_{\mathrm{t}\beta}$), see the appendix for
further details on the QPC detector response.

For a quantum dot with multiple charge states we have to require that the
expansion of the scattering amplitudes $t_n = \sqrt{T_n} e^{i\theta_n}$ and
$r_n = \sqrt{1-T_n}e^{i\chi_n}$ of the QPC detector scale linearly in the
charge $Q_n$ of the dot, i.e., $T_n=T-Q_n \delta T$, $\theta_n=\theta+Q_n
\delta \theta$, and $\chi_n=\chi+ Q_n \delta \chi$. A straightforward
calculation then shows that the results (\ref{eq:det_R}) and (\ref{eq:det_I})
for the detector response functions as well as the final results
(\ref{eq:RS_c}) and (\ref{eq:IS_c}) remain unchanged.

\subsection{Strong measurement}\label{sec:charge_sc}

When performing a strong measurement of a quantum dot with a binary charge it
is sufficient to invoke individual electron pulses as quantum memories. For a
strong dot--QPC coupling, we require a one-to-one relation between the
presence of a charge on the dot and the outcome of the measurement, i.e.,
$|\phi_0\rangle = |\varphi_\mathrm{t}\rangle$ and $|\phi_1\rangle =
|\varphi_\mathrm{r}\rangle$, see Eq.\ (\ref{eq:kroneckerscattering}), or
$s^\mathrm{r}_1 = 1$ and $s^\mathrm{r}_0 = 0$. This is achieved by tuning the
QPC such as to generate a unique scattering outcome with $|r_1|=1$, $|r_0|=0$
and $|t_0|=1$, $|t_1|=0$, i.e., the presence of a charge $Q_1 = 1$ on the dot
reflects the QPC electron back to the left. In this case, it is the reflection
probability $P_\mathrm{rr}(\tau_{21})$ that directly traces the charge $\Q$
and according to (\ref{eq:Pab}), we have to determine the expression
\begin{equation} \label{eq:Prr}
   P_\mathrm{rr}(\tau_{21})
   = \sum_m \Bigl| \sum_n r_m U_{mn}(\tau_{21})
                          r_n \psi_n(\tau_1) \Bigr|^2.
\end{equation}
With $r_m = \delta_{m1}$, we obtain the simple result $P_\mathrm{rr}(\tau_{21})
= |U_{11}(\tau_{21})\psi_1|^2$ (see also (\ref{eq:Panbm}) with $n,m = 1$ and
$\alpha_1 = \beta_1 = \mathrm{r}$) and since $Q_n = \delta_{n1}$, we find the
projected correlator (see Eq.\ (\ref{eq:SPOO}))
\begin{align}\label{eq:SPQQ}
   S_{\Qi\Qi}^P(\tau_1,\tau_2) = 
        P_\mathrm{rr}(\tau_{21}) = |U_{11}(\tau_{21})\psi_1|^2.
\end{align}
Similarly, the probability to find no charge on the dot in either of the two
measurements is $P_\mathrm{tt} = |U_{00}\psi_0|^2$, while the mixed results
are $P_\mathrm{tr} = |U_{10}\psi_0|^2$ and $P_\mathrm{rt} = |U_{01}\psi_1|^2$.

The strong measurement of the charge correlator for a multi-charge quantum dot
quite naturally involves trains of electron pulses \cite{averin:05}, with the
number $J$ of electrons in each train sufficiently large to distinguish the
different charge eigenvalues $Q_n$ of the dot. The separation $\delta\tau$
between electron pulses within a train has to be sufficiently long in order to
allow for their separate detection (i.e., counting), while the train duration
$J\, \delta \tau$ must remain small on the scale $\tau_\mathrm{sys}$ of the
dot's dynamics.

When going over from qubit registers to electron trains scattered at the QPC
we replace the `magnetization' (\ref{eq:Mn}) by the disbalance between reflected
and transmitted electrons,
\begin{align}\label{eq:Dn}
   D(n) = \sum_\mu  \mu [P_\mathrm{r}^\mu(n) - P_\mathrm{t}^\mu(n)] 
   =J[R-T +2 Q_n \delta R],
\end{align}
where we have assumed a linear QPC characteristic with a reflection
probability scaling linearly with the dot's charge $Q_n$, $R_n = R + Q_n
\delta R$. Operating the QPC at the symmetry point $T = R = 1/2$, we can
determine the combination $S_\mathrm{rr} - S_\mathrm{rt} - S_\mathrm{tr} +
S_\mathrm{tt}$ and relate this quantity to the projected charge correlator
$S_{QQ}^P (\tau_{21})$,
\begin{align}\label{eq:SQQP}
   S_{QQ}^P(\tau_{21})
            = \frac{S_\mathrm{rr}-S_\mathrm{rt}-S_\mathrm{tr}+S_\mathrm{tt}}
                   {4 J^2 \delta R^2}.
\end{align}
Operating the QPC away from the symmetric point, one has to determine the
weighted sum $T^2 S_\mathrm{rr} - TR\,S_\mathrm{rt} - RT\,S_\mathrm{tr} + R^2
S_\mathrm{tt}$ instead and divide by $J^2 \delta R^2$ to arrive at the projected
correlator $S_{QQ}^P$.

\subsection{Finite-width memory wave-packets}\label{sec:finitewidth}

Above we have assumed an instantaneous (within a short time $\delta\tau$)
entanglement between the system and the memory states, requiring that both the
width $\tau_{\rm wp}$ of the wave-packet and the scattering time $\tau_{\rm
sca}$ at the QPC satisfy $\tau_{\rm wp},\tau_{\rm sca}\ll \tau_{\rm sys}$,
where $\tau_{\rm sys}$ denotes the characteristic time scale of the system.
Here, we allow for a spread in time of the detector's electron wave function
and drop the condition $\tau_{\rm wp}\ll \tau_{\rm sys}$, i.e., we assume that
$\tau_{\rm wp}\lesssim \tau_{\rm sys}$ while the scattering event itself
remains fast, $\tau_{\rm sca}\ll \tau_{\rm sys}$.  In general terms, this
corresponds to a measurement which probes the system sharply ($\tau_{\rm
sca}\ll \tau_{\rm sys}$) during some finite time ($\tau_{\rm wp}\lesssim
\tau_{\rm sys}$; longer measurement times $\tau_{\rm wp} > \tau_{\rm sys}$ do
not provide meaningful results).

Let us suppose that the $j$-th wave-packet incident on the QPC around $\tau_j$
is described by the wave function $f^{\scriptscriptstyle (j)}(\tau)$ which is
normalized ($\int d\tau |f^{\scriptscriptstyle (j)}(\tau)|^2=1$) and peaked at
the time $\tau_j$. Assuming instantaneous scattering, we obtain the final
state
\begin{align}
   &|\tilde{\Psi}_f\rangle 
   = \int d\tau_1'\, f^{\scriptscriptstyle (1)}(\tau_1')\int d\tau_2' 
                   \,f^{\scriptscriptstyle (2)}(\tau_2')\\
      &\times\sum_{l,m,n} U_{lm}(\tau_{f2}') U_{mn}(\tau_{21}') 
      \psi_n(\tau_1')\, |l\rangle
      \, |\phi^{\scriptscriptstyle (1)}_{n}(\tau_1') \rangle
      \, |\phi^{\scriptscriptstyle (2)}_{m}(\tau_2') \rangle,
      \nonumber
\end{align}
with $\tau_{f2}'=\tau_f-\tau_2'$ and $\tau_{21}'=\tau_2'-\tau_1'$ and the
outgoing memory states $|\phi^{\scriptscriptstyle (j)}_{n}(\tau_j') \rangle$
that have been scattered at the QPC at time $\tau_j'$. Making use of $\langle
\phi^{\scriptscriptstyle (j)}_{n'}(\tau_j'') |\phi^{\scriptscriptstyle
(j)}_{n}(\tau_j') \rangle = \delta(\tau_j'-\tau_j'')\langle
\phi^{\scriptscriptstyle (j)}_{n'} |\phi^{\scriptscriptstyle (j)}_{n}\rangle$,
we obtain the smeared probabilities
\begin{align}
   \tilde{P}_{\alpha\beta}(\tau_{21})
   = \!\!\int\!\! d\tau_1' \!\!\int\!\! d\tau_2'\, |f^{\scriptscriptstyle (1)}(\tau_1')|^2 
   |f^{\scriptscriptstyle (2)}(\tau_2')|^2 
   P_{\alpha\beta}(\tau_{21}')
   \label{eq:Pabfinitewavepacket}
\end{align}
with $P_{\alpha\beta}$ given by Eq.\ \eqref{eq:Pab}. The finite width wave-packets enter
as integration kernels.  While for sharp wave-packets, the entanglement
between system and memories (i.e., the `measurement') takes place at times
$\tau_1$ and $\tau_2$, for broader wave-packets, the entanglement arises
within a finite time $\tau_{\rm wp}$ around $\tau_1$ and $\tau_2$ with
distribution functions $|f^{\scriptscriptstyle (1)}(\tau_1')|^2$ and
$|f^{\scriptscriptstyle (2)}(\tau_2')|^2$. As a consequence, for the case of
strong coupling where the entanglement gives rise to projective measurements,
the times of projection are not fixed but distributed with the distribution
functions above.

Note that the result \eqref{eq:Pabfinitewavepacket} is only valid for
negligible scattering time $\tau_{\rm sca}$ in the QPC.  If the scattering
time $\tau_{\rm sca}$ is finite compared to the system time $\tau_{\rm sys}$,
the effect of interaction cannot be accounted for by the scattering matrices
$\hat{s}_n$ of the QPC depending on the system state, but the interaction 
during the scattering has to be treated in more detail.

The limit of $\tau_{\rm sca} \ll \tau_{\rm sys}$ considered above has also
implications for the resulting back-action: While the back-action for
$\tau_{\rm sca} \ll \tau_{\rm sys}$ only consists of dephasing, i.e.,
a suppression of off-diagonal elements of the system's density matrix, for
finite $\tau_{\rm sca}$, the back-action of the measurement on the system goes
beyond pure dephasing and alters the system's dynamics.

\subsection{Higher-order correlators}\label{sec:higher}

The study of higher-order correlators is straightforward, e.g., to measure a
third-order charge correlator, we send three electrons scattering from the QPC
at times $\tau_1<\tau_2<\tau_3$ and obtain the probabilities
\begin{equation} \label{eq:Pabc}
   P_{\alpha\beta\gamma} 
   = \sum_l \Bigl| \sum_{m,n} s_l^\gamma U_{lm}(\tau_{32})
   					    s_m^\beta U_{mn}(\tau_{21}) 
                                   	    s_n^\alpha \psi_n(\tau_1) \Bigr|^2
\end{equation}
describing electrons transmitted across ($\alpha,\beta,\gamma=\mathrm{t}$) or
reflected from ($\alpha,\beta,\gamma=\mathrm{r}$) the QPC. For weak coupling,
its irreducible part can be recast in the form
\begin{align}\label{eq:Pabcirr}
   P_{\alpha\beta\gamma}^{\rm irr}
   = \sum_{\sigma \sigma'=\pm}
       S^{\scriptscriptstyle (1),\bar{\sigma}}_{\mathrm{det},\alpha}
       S^{\scriptscriptstyle (2),\bar{\sigma}'}_{\mathrm{det},\beta}
       S^{\scriptscriptstyle (3),-}_{\mathrm{det},\gamma}\,
        S^{\sigma\sigma',\mathrm{irr}}_{QQQ}(\tau_1,\tau_2,\tau_3)
\end{align}
with $\bar{\sigma}=-\sigma$, the detector responses 
$S^{\scriptscriptstyle (j),+}_{\mathrm{det},\alpha}
=\mathcal{R}S^{\scriptscriptstyle (j)}_{\mathrm{det},\alpha}$ and
$S^{\scriptscriptstyle (j),-}_{\mathrm{det},\alpha}
=\mathcal{I}S^{\scriptscriptstyle (j)}_{\mathrm{det},\alpha}$, and
the third-order correlators
\begin{align}
S^{\sigma\sigma',{\rm irr}}_{QQQ}
	=c_\sigma c_{\sigma'}
		\llangle [\Q(\tau_1), [\Q(\tau_2),
                \Q(\tau_3)]_{\sigma'}]_{\sigma}\rrangle
\end{align}
with the constants $c_+=1/2$ and $c_-=-i$ and $[\cdot,\cdot]_-=[\cdot,\cdot]$
resp. $[\cdot,\cdot]_+=\{\cdot,\cdot\}$ (note that (anti-)symmetrized charges
in $S^\mathrm{irr}_{QQQ}$ (encoded in $\sigma$) relate to opposite detector
response functions (encoded by $\bar\sigma$).  This result agrees with the one
in Ref.~\onlinecite{bayandin:08} obtained with the help of the von Neumann
projection postulate and shows that only {\it Keldysh time-ordered charge
correlators} are measurable.

\subsection{Experimental implementation}\label{sec:charge_exp}

Our general concept of deferred repeated measurements has been formulated with
quantum memories, e.g., qubits, qubit registers, and qudits, see Sec.\
\ref{sec:uni}. In our application of these general considerations, the
measurement of a charge correlator with the help of a quantum point contact
(Sec.\ \ref{sec:charge}), the quantum memories have been replaced by scattered
electrons (flying qubits). It then is natural to seek for an experimental
implementation, where the final measurement of the quantum memories, i.e., the
scattered electrons, can be cast into a measurement of currents and noise
rather than an individual detection of qubit states. Such an implementation is
proposed below.

\begin{figure}
\begin{center}
\includegraphics[width=8cm]{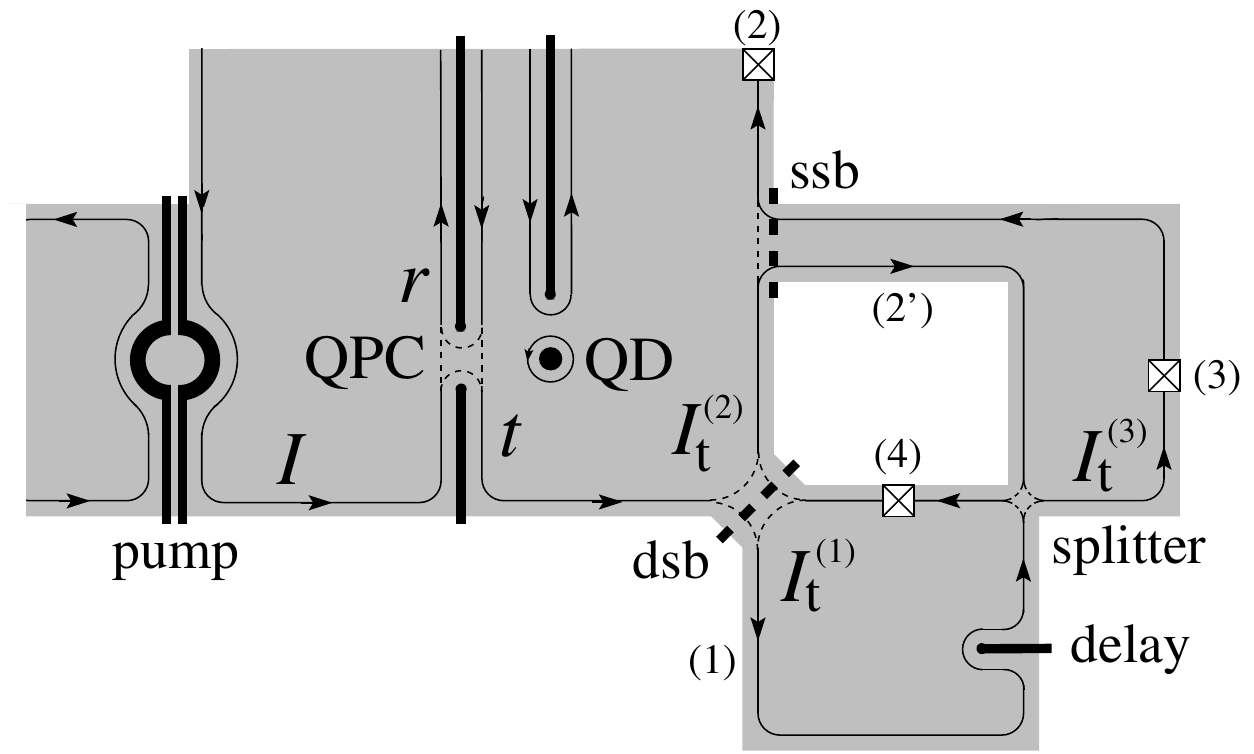}
\end{center}
\caption{\label{fig:exp} Schematic illustration of a quantum Hall device with
a single electron pump (left), the QPC--QD setup (middle) and a detector
arrangement (right), inspired by Refs.\
\onlinecite{fletcher:13,bocquillon:13}.  Pairs of electrons are injected by
the single electron pump with a time delay $\tau_{21}$ and scattered by the
quantum point contact with scattering coefficients $r$ and $t$ depending on
the charge state of the quantum dot.  The dynamically switchable
barrier\cite{fletcher:13} (dsb) separates the first and second electron (flying
qubits $j=1$ and $j=2$) into arms 1 and 2. Closing the static switchable
barrier (ssb) and measuring the currents $I_\mathrm{t}^{\scriptscriptstyle (1)} =
I_\mathrm{t}^{\scriptscriptstyle (3)} +I_\mathrm{t}^{\scriptscriptstyle (4)}$
and $I_\mathrm{t}^{\scriptscriptstyle (2)}$ provides the anti-symmetrized
charge-charge correlator $\mathcal{I} S^\mathrm{irr}_{QQ}$. Opening the static
switchable barrier (ssb), having the two particle streams along the trajectories
1 and 2' interfere at the 50/50 beam splitter, and measuring the current noise
$S^{\scriptscriptstyle(3)}(\tau)$ as a function of the tunable delay $\tau$ in
the detector 3 provides the symmetrized correlator $\mathcal{R}
S^\mathrm{irr}_{QQ}$.}
\end{figure}

By now, several experiments have demonstrated the controlled generation of
individual electron pulses \cite{feve:07,fletcher:13,bocquillon:13} and even
the detection of such pulses via qubit detectors seems to be in reach
\cite{thalineau:14}.  A setup particularly well suited within the current
context is the one of Fletcher {\it et al.} \cite{fletcher:13}, involving a
single electron pump and a time-correlated detector setup in a quantum Hall
device, see Fig.\ \ref{fig:exp} for an illustration. Pairs of electron
pulses of width $\sim 100$ ps are injected at a typical rate $\nu_p \sim (10
~\mathrm{ns})^{-1}$ and are well suited to probe the dynamics of a dot with a
system time $\tau_\mathrm{sys} \sim 100 - 1000$ ns.  The detector involves a
dynamically switchable barrier (dsb in Fig.\ \ref{fig:exp}) that can be
switched on or off with a nanosecond precision in time.

We propose an experiment implementing a weak measurement of a charge
correlator along the lines of Sec.\ \ref{sec:charge_wc}, where electron pairs
with a tunable delay time in the range $\tau_{21} = 10 - 100$ nanoseconds are
injected by the single electron pump (with injection currents in the $I =
2e\nu_p \sim 10$ pA regime) and scattered at the QPC probing the quantum dot
QD.  The transmitted electrons arrive at the detector dsb after a delay time
$\tau_d$.  The detector barrier dsb is switched on at a time $\tau_d +
\tau_{21}/2$ in order to let the first electron pass along the path $1$ and
deflect the second electron along $2$, see figure \ref{fig:exp}. With the
static barrier (ssb) closed, the measured currents
$I_\mathrm{t}^{\scriptscriptstyle (1)}$ and $I_\mathrm{t}^{\scriptscriptstyle
(2)}$ resulting from the transmission of the first and second electron,
respectively, can be related to the probabilities $P_{\alpha\beta}$ via
\begin{align}\label{eq:I1}
   I_\mathrm{t}^{\scriptscriptstyle (1)}/e\nu_p
   = P_\mathrm{tt} + P_\mathrm{tr}, \\
   \label{eq:I2}
   I_\mathrm{t}^{\scriptscriptstyle (2)}/e\nu_p
   =  P_\mathrm{tt} + P_\mathrm{rt}.
\end{align}
The two sums $\sum_{\beta} P_{\alpha\beta}$ and $\sum_{\alpha}
P_{\alpha\beta}$ in the above equations are fundamentally different in a
subtle way: summing over the second particle $\beta$ produces the trivial
probability $P_\alpha^{\scriptscriptstyle (1)} = \langle \Psi_f
|\hat{p}^{\scriptscriptstyle (1)}_\alpha| \Psi_f\rangle$ for the (independent)
transmission ($\alpha = \mathrm{t}$) or reflection ($\alpha = \mathrm{r}$) of
the first electron by the QPC, $\sum_\beta P_{\alpha\beta} =
P_\alpha^{\scriptscriptstyle (1)}$. Hence, the first equation (\ref{eq:I1})
reduces to $I_\mathrm{t}^{\scriptscriptstyle (1)}/e\nu_p =
P_\mathrm{t}^{\scriptscriptstyle (1)}$ and thus contains only information
about the mean charge on the dot.

On the contrary, summing over the first particle $\alpha$ {\it does not}
generate $P_\beta^{\scriptscriptstyle (2)}$ as the interaction of the first
electron at the earlier time $\tau_1$ introduces nontrivial correlations, see
Eq.\ (\ref{eq:Pab}). This becomes apparent when going over to irreducible
probabilities $P_{\alpha\beta}^\mathrm{irr} = P_{\alpha\beta} -
P_\alpha^{\scriptscriptstyle (1)} P_\beta^{\scriptscriptstyle (2)}$ and using
the normalization $\sum_\alpha P_\alpha^{\scriptscriptstyle (j)} = 1$.  Then,
the second equation (\ref{eq:I2}) becomes $I_\mathrm{t}^{\scriptscriptstyle
(2)}/e\nu_p = P_\mathrm{tt}^\mathrm{irr} + P_\mathrm{rt}^\mathrm{irr} +
P_\mathrm{t}^{\scriptscriptstyle (2)}$ which includes information about the
dot's charge correlator. Assuming a time-independent mean charge on the dot,
we have $P_\mathrm{t}^{\scriptscriptstyle (2)} =
P_\mathrm{t}^{\scriptscriptstyle (1)} = I_\mathrm{t}^{\scriptscriptstyle
(1)}/e\nu_p$ and using the general result (\ref{eq:Pabweak}), the measured
currents are easily transformed to provide the antisymmetrized correlator
\begin{align}\label{eq:IS_exp}
   \mathcal{I} S^\mathrm{irr}_{QQ}
   = \frac{(I_\mathrm{t}^{\scriptscriptstyle (2)}
   - I_\mathrm{t}^{\scriptscriptstyle (1)})/e\nu_p}
   {\delta T (R \delta\chi + T \delta\theta)}.
\end{align}
Note that the evaluation of the irreducible probability
$P_\mathrm{tt}^\mathrm{irr} + P_\mathrm{tr}^\mathrm{irr}$ with the help of
(\ref{eq:Pabweak}) indeed provides a vanishing result,
$P_\mathrm{tt}^\mathrm{irr} + P_\mathrm{tr}^\mathrm{irr}  \propto
(\mathcal{I}S_\mathrm{det,r} + \mathcal{I}S_\mathrm{det,t}) = 0$, explicitly
demonstrating that the sum $P_\mathrm{tt}^\mathrm{irr} +
P_\mathrm{tr}^\mathrm{irr}$ contains no correlations.

In order to find the symmetric correlator $\mathcal{R}S^\mathrm{irr}_{QQ}$ one
has to measure a time correlator on the transmitted channel. This can be
conveniently done with the help of a Hong-Ou-Mandel type splitter as
implemented in the experiment of Bocquillon {\it et al.} \cite{bocquillon:13}
and sketched in Fig.\ \ref{fig:exp}.  In this experiment, the dynamically
switchable barrier (dsb) again splits the two electrons in each pair to
propagate along the paths 1 and 2 $\rightarrow 2'$, respectively.  The static
barrier (ssb) is left open, such that the two particles interfere in the
splitter. Measuring the current noise $S^{\scriptscriptstyle (3)}(\tau)$ in
channel 3 as a function of mutual delay $\tau$ (tuned via an additional gate
in loop 1, see figure) then provides all information needed to construct
$\mathcal{R}S^\mathrm{irr}_{QQ}$.

We can calculate the evolution of the state through the Hong-Ou-Mandel setup
using the wave function $|\Psi_{f}\rangle = |\Psi_{f}\rangle_\mathrm{tt} +
|\Psi_{f}\rangle_\mathrm{tr} + |\Psi_{f}\rangle_\mathrm{rt} +
|\Psi_{f}\rangle_\mathrm{rr}$ describing the two-electron state after the
scattering events at the QPC, with $P_{\alpha\beta} =
\,_\mathrm{\alpha\beta}\langle \Psi_f| \Psi_f\rangle_\mathrm{\alpha\beta}$,
i.e., the individual components are not normalized and only $\langle \Psi_f|
\Psi_f\rangle =1$.  Due to the orthogonality of these four components, the
particle numbers $\hat{N}^{\scriptscriptstyle (i)}$, $i = 3,4$, emerging from
our Hong-Ou-Mandel splitter can be analyzed term by term.  In particular, the
particle number fluctuations $\langle \Psi_f | (\delta \hat{N}_3)^2 | \Psi_f
\rangle$ in channel 3 involve single-particle and two-particle contributions,
$_\mathrm{rt}\langle \Psi_f | \hat{N}_3 | \Psi_f \rangle_\mathrm{rt} = (1/2)
\cdot 1 \cdot P_\mathrm{rt}$ and $_\mathrm{rt}\langle \Psi_f |
\hat{N}_3^2 | \Psi_f \rangle_\mathrm{rt} = (1/2) \cdot 1^2 \cdot P_\mathrm{rt}$,
hence,
\begin{equation}\label{eq:dN32}
   _\mathrm{rt}\langle \Psi_f |(\delta \hat{N}_3)^2| \Psi_f \rangle_\mathrm{rt}
  = \frac{1}{4} \cdot P_\mathrm{rt},
\end{equation}
and the same result holds true for the $| \Psi_f \rangle_\mathrm{tr}$
scattering component. While there is no contribution from
$|\Psi_{f}\rangle_\mathrm{rr}$, the one originating from
$|\Psi_{f}\rangle_\mathrm{tt}$ depends on the time delay $\tau$. Let $f_1(x)$
and $f_2(x)$ denote the two electron wave-packets propagating along the
incoming paths $1$ and $2$ of the HOM-interferometer. As shown in
Ref.~\onlinecite{lebedev:08}, we can relate the number fluctuation in the
component $|\Psi_{f}\rangle_\mathrm{tt}$ to the overlap of wavefunctions as
\begin{equation}
   _\mathrm{tt}\langle\Psi_f| (\delta \hat{N}_3)^2 |\Psi_f\rangle_\mathrm{tt}
   = \frac12\Bigl( 1 - |\langle f_1 | f_2\rangle|^2\Bigr) P_\mathrm{tt},
\end{equation}
where $\langle f_1|f_2\rangle = \int dx f_1^*(x) f_2(x)$.  For a large time
delay between the first and second electron, these do not interfere, $\langle
f_1 | f_2\rangle =0$, such that the fluctuations are just the double of
(\ref{eq:dN32}),
\begin{align}
  _\mathrm{tt}\langle \Psi_f | (\delta \hat{N}_3)^2 | \Psi_f \rangle_\mathrm{tt}
   = \frac{1}{2}\, P_\mathrm{tt}.
\end{align}
Compensating the original time delay $\tau=-\tau_{21}$ (such that the two
electrons arrive simultaneously at the splitter), we have
\begin{align}
  _\mathrm{tt}\langle \Psi_f | (\delta \hat{N}_3)^2 | \Psi_f \rangle_\mathrm{tt}=0,
\end{align}
since the Pauli exclusion forces the two particles to propagate to different
channels.  The final result $\langle \Psi_f| (\delta\hat{N}_3)^2
|\Psi_f\rangle$ then involves separately the probabilities $P_\mathrm{rt}
+P_\mathrm{tr}$ and $P_\mathrm{tt}$,
\begin{align}
\langle \Psi_f| (\delta \hat{N}_3)^2 |\Psi_f \rangle_{\infty}
&=
\frac{1}{4}(P_\mathrm{rt} +P_\mathrm{tr}) +\frac{1}{2}P_\mathrm{tt},
\end{align}
when the delay between electron pulses is not compensated and
\begin{align}
\langle \Psi_f| (\delta \hat{N}_3)^2 |\Psi_f\rangle_0
&=
\frac{1}{4}(P_\mathrm{rt} +P_\mathrm{tr}),
\end{align}
when the time delay is properly compensated, $\tau=-\tau_{21}$.

In a next step, we express the charge fluctuations in channel $3$ through the
irreducible current-current correlator,\cite{lebedev:08}
\begin{equation}
   \langle\Psi_f| (\delta \hat{N}_3)^2|\Psi_f\rangle
   = \frac1{e^2} \int_W dt_1 dt_2 \, 
   \langle\langle \hat{I}^{\scriptscriptstyle (3)}(t_1)
   \hat{I}^{\scriptscriptstyle (3)}(t_2)\rangle\rangle,
   \label{eq:II3}
\end{equation}
where the time-window $W$ is centered around the arrival time of the
wave-packets at the detector $3$ and has a width of the order of $\Delta\tau =
\nu_p^{-1}$, with $\nu_p$ the rate of pair injection by the pump.  
The particle number fluctuations then can be expressed through the
low-frequency current noise $S^{\scriptscriptstyle (3)}(\omega)$,
\begin{equation}
   \langle \Psi_f|(\delta\hat{N}_3)^2|\Psi_f\rangle
    = \int \frac{d\omega}{2\pi}\, \frac{S^{\scriptscriptstyle (3)}(\omega)}{e^2} \,
   \frac{\sin^2(\omega/ 2\nu_p)}{(\omega/2)^2},
\end{equation}
with $\omega \leq \nu_p$. Neglecting the frequency dependence of the noise at
small $\omega$, $S^{\scriptscriptstyle (3)}(\omega) \sim S^{\scriptscriptstyle
(3)}(0)$, we obtain
\begin{equation}
   \langle \Psi_f| (\delta\hat{N}_3)^2|\Psi_f\rangle = S^{\scriptscriptstyle
   (3)}/e^2 \nu_p,
\end{equation}
and hence
\begin{equation} \label{eq:PrtS}
   P_\mathrm{rt} + P_\mathrm{tr}= 4\frac{S^{\scriptscriptstyle (3)}_0}{e^2 \nu_p}, \quad
   P_\mathrm{tt} = 2\frac{S^{\scriptscriptstyle (3)}_\infty
   -S^{\scriptscriptstyle (3)}_0}{e^2 \nu_p}.
\end{equation}

Using Eq.\ \eqref{eq:Pabweak}, we can derive an alternative (more symmetrized)
version of Eq.\ \eqref{eq:RS_c} which involves just the combinations
$P_\mathrm{rt}^\mathrm{irr} + P_\mathrm{tr}^\mathrm{irr}$ and
$P_\mathrm{tt}^\mathrm{irr}$,
\begin{align}\label{eq:RSQQs}
  &\mathcal{R} S^\mathrm{irr}_{QQ} \\ \nonumber
  &~~~= \frac{(R\delta\chi)^2\,
  P^\mathrm{irr}_\mathrm{tt} - T\delta\theta\,R\delta\chi\,
 (P^\mathrm{irr}_\mathrm{rt} + P^\mathrm{irr}_\mathrm{tr})
  + (T\delta\theta)^2 \, P^\mathrm{irr}_\mathrm{rr}}
  {(\delta T)^2 (R\delta\chi + T \delta\theta)^2}.
\end{align}
Defining the detector parameters $\kappa =T\delta \theta/(R\delta \chi + T
\delta\theta)$ and going over to irreducible probabilities (with
$P_\mathrm{t}^{\scriptscriptstyle (1)}=P_\mathrm{t}^{\scriptscriptstyle
(2)}=I_{t}^{\scriptscriptstyle (1)}/e\nu_p$ independent of time), we obtain
the final result
\begin{eqnarray}\label{eq:RSQQ_exp}
  &&(\delta T)^2\, \mathcal{R} S^\mathrm{irr}_{QQ} =
  \\
  &&=
  2(1\!-\!2\kappa) \frac{S^{\scriptscriptstyle (3)}_\infty
  - S^{\scriptscriptstyle (3)}_0}{e^2\nu_p}
   + 4\kappa \frac{S^{\scriptscriptstyle (3)}_0}{e^2\nu_p}
   + \kappa^2 - \Bigl( \frac{I_{t}^{\scriptscriptstyle (1)}}{e\nu_p} 
   - \kappa\Bigr)^2.
  \nonumber
\end{eqnarray}

Note that in this setup, the final measurement of qubit memories does not
require fast or time resolved detection schemes, but merely relies on the
measurement of average currents and low-frequency noise. This is due to the
fact, that all timing tasks are realized by the properly time-delayed electron
pulses in the incoming channel and the dynamically switchable gate (dsb) which
separates the electron pairs; both elements have been realized in an
experiment\cite{fletcher:13}. Hence, the new measurement scheme, combined with
novel elements from electron quantum optics, allows to shift the (difficult)
timing issues in the measurement of a time-correlator from the detector to the
source.

A {\it strong coupling} between the quantum dot and the quantum point contact
provides us with a projected correlator. For the simplest case of a binary
charge on the dot with values $Q=0,1$, the projected charge correlator is
given by Eq.\ \eqref{eq:SPQQ}, $S_{QQ}^P = P_\mathrm{rr}$. Making use of the
normalization $P_\mathrm{tt} + P_\mathrm{rt} + P_\mathrm{tr} + P_\mathrm{rr} =
1$ and the result Eq.\ (\ref{eq:PrtS}), we find that
\begin{align}\label{eq:SPQQexp}
   S^P_{QQ}=1-2\,\frac{S^{\scriptscriptstyle (3)}_0 + S^{\scriptscriptstyle
   (3)}_\infty}{e^2\nu_p}.
\end{align}

\section{Conclusion}\label{sec:con}

In conclusion, we have applied the principle of deferred measurement to the
problem of repeated measurement and have derived physical expressions for the
two- and multi-time correlators. The measurement involves the inclusion of
quantum memories which are entangled with the system at specific times
$\tau_j$ where the system observable is to be probed. The expanded system plus
memories undergoes a unitary evolution until the very end, where the result is
extracted via application of Born's rule to the memories. The measured
probabilities $P_{\alpha\beta}$ (see Eq.\ (\ref{eq:Pab})) or memory
correlators $S_{\alpha\beta}$ (see Eq.\ (\ref{eq:Sab})) then can be combined
to extract the desired system correlators. The limits of weak and strong
measurements provide the standard (anti-) symmetrized and projected time
correlators previously obtained by invoking the (non-unitary) von Neumann
projection. The general results have been illustrated by using qubits and
qubit registers as quantum memories. Our analysis sheds new light on the
problem of repeated measurement and illustrates the usefulness of qubits as
sensitive measurement devices.

Although our paper's main results are rather on the conceptual side, one could
imagine an implementation of such a deferred measurement in an experiment.  A
system that naturally lends itself for a realization of these ideas is the
classic mesoscopic setup which probes the charge of a quantum dot through a
quantum point contact. The individual scattered electrons in the QPC can be
understood as flying qubits which are either transmitted or reflected, with
amplitudes depending on the charge state of the quantum dot.  In particular,
the qubit register required in the strong measurement of a dot with a
multi-valued charge is easily implemented in terms of finite trains of
electrons. We have applied our formalism to this situation and derived the
corresponding expressions for a weak and strong measurement.

The experimental implementation of these ideas requires a system control that
can only be met with a modern quantum engineering approach. Recent
developments in electron quantum optics provide controlled single electron
pulses and allow for their time resolved manipulation/detection on a
sub-nanosecond time scale \cite{bocquillon:13,fletcher:13}. Using the setup of
Ref.\ \onlinecite{fletcher:13} as a base and augmenting it with a
Hong-Ou-Mandel type analyzer \cite{bocquillon:13}, we propose to include
a quantum point contact and a quantum dot in order to realize the principle of
repeated measurement by a deferred measurement of quantum memories.

As a final remark, one may appreciate the relation of the deferred measurement
principle to Everett's idea of a multiverse \cite{tegmark:07,dewitt:73}.
Rather than applying a projection after the first measurement and pursuing a
single further evolution (of the system = `universe'), the principle of a
deferred measurement involving the system's entanglement with a quantum memory
enhances the overall dimensionality, e.g., for a qubit memory the
dimensionality is doubled (with two `universes' evolving in parallel). It is
then only the final measurement which determines which evolution (i.e., which
`universe') actually has been realized.

\begin{acknowledgements}

We thank Renato Renner for discussions and acknowledge financial support from
the Swiss National Science Foundation through the National Center of
Competence in Research on Quantum Science and Technology (QSIT), the Pauli
Center for Theoretical Studies at ETH Zurich, and the RFBR Grant No.\
14-02-01287.

\end{acknowledgements}

\appendix*
\section{Detector properties}\label{sec:det}
A quick overview is provided by the example of a {\it $\delta$-function
scatterer:} Expressing the strength of the scatterer $\hbar^2 \lambda /m $ for
the two charge states by $\lambda_0 =\lambda$ and $\lambda_1=\lambda+\delta
\lambda$, an incoming state with wave vector $k$ is transmitted with amplitude
$t_n=k/(k+i\lambda_n)$, $n \in \{0,1\}$. Expanding the transmission
$T_n=k^2/(k^2+\lambda_n^2)$ and phase $\theta_n=-\arctan(\lambda_n/k)$, we
find the modifications $\delta T$ and $\delta\theta = \delta\chi$ (for a
symmetric scatterer) in the scattering characteristic of the QPC upon charging
the dot
\begin{align}
   \delta T&\approx-\frac{2k^2\lambda^2}{(k^2+\lambda^2)^2}
    \frac{\delta \lambda}{\lambda},\\
   \delta \theta = \delta \chi &=-\frac{k\lambda}{k^2+\lambda^2}
     \frac{\delta \lambda}{\lambda}.
\end{align}
In the limit of a large incoming energy, i.e., $k\gg\lambda$, we find $\delta
T\approx-2\lambda \,\delta \lambda/k^2$ and $\delta \theta=\delta
\chi\approx-\delta \lambda/k$ and hence $\delta\theta,\delta\chi\gg \delta T$;
with $T\approx 1$ and $R\approx \lambda^2/k^2\ll 1$, we have $T|\delta\theta|
\gg |\delta T| \gg R |\delta \chi|$ and therefore $|\mathcal{I}S_\mathrm{det,t}|
\gg |\mathcal{R}S_\mathrm{det,r/t}|\gg |\mathcal{I}S_\mathrm{det,r}|$.  For a
small incoming energy $k\ll\lambda$, we obtain $\delta T \approx -2k^2 \delta
\lambda/\lambda^3$ and $\delta \theta \approx -k \delta \lambda/\lambda^2$ and
using $T\approx k^2/\lambda^2$ and $R\approx 1$, we find $|\mathcal{I}
S_\mathrm{det,r}|\gg |\mathcal{R} S_\mathrm{det,r/t}|\gg |\mathcal{I}
S_\mathrm{det,t}|$.  When $k\approx\lambda$, all response functions are of the
same order.

Alternatively, we can consider a {\it single electron transistor (SET)} with
the level position $k_{\mathrm{res},n}$ affected by the capacitive coupling
and depending on the dot's charge state $|n\rangle$, i.e., $k_{\mathrm{res},0}
= k_{\mathrm{res}}$ and $k_{\mathrm{res},1}= k_{\mathrm{res}}+\delta
k_\mathrm{res}$. The transmission coefficient is given by $t_n = i\gamma /
(k-k_{{\rm res},n} + i\gamma)$, where $\gamma$ is the level width. Again expanding 
$T_n=\gamma^2/((k-k_{{\rm res},n})^2+\gamma^2)$ and $\tan
\theta_n=(k-k_{{\rm res},n})/\gamma$ for small $\delta k_{\rm res}\ll k_{\rm res}$,
we find
\begin{align}
\delta T&=-\frac{2(k-k_{\rm res})\gamma^2 
\delta k_{\rm res}}{[(k-k_{\rm res})^2+\gamma^2]^2},\\
\delta\theta = \delta \chi &=-\frac{\gamma \delta k_{\rm res}}
                                   {(k-k_{\rm res})^2+\gamma^2}.
\end{align}
For incoming electrons on resonance with the level, i.e., $|k-k_{\rm res}|\ll
\gamma$, we obtain $\delta T\approx -2 (k-k_{\rm res})\delta k_{\rm
res}/\gamma^2$ and $\delta \theta=\delta \chi \approx - \delta k_{\rm
res}/\gamma $, such that  $\delta\theta,\delta\chi\gg \delta T$ and using
$T\approx 1$ and $R\approx (k-k_{\rm res})^2/ \gamma^2$, we find that
$|\mathcal{I}S_\mathrm{det,t}|\gg |\mathcal{R}S_\mathrm{det,r/t}|\gg
|\mathcal{I}S_\mathrm{det,r}|$.  On the other hand, for off-resonant electrons
$\delta T\approx -2 \gamma^2 \delta k_{\rm res}/(k-k_{\rm res})^3$ and $\delta
\theta=\delta \chi\approx - \gamma \delta k_{\rm res}/(k-k_{\rm res})^2$, such
that $\delta\theta,\delta\chi\gg \delta T$ and using $T\approx
\gamma^2/(k-k_{\rm res})^2$ and $R\approx1$, we find
$|\mathcal{I}S_\mathrm{det,r}|\gg |\mathcal{R}S_\mathrm{det,r/t}|\gg
|\mathcal{I}S_\mathrm{det,t}|$.  When $|k-k_\mathrm{res}|\approx\gamma$, all
response functions are of the same order.

\begin{figure}[htbp]
\begin{center}
\includegraphics[scale=.66]{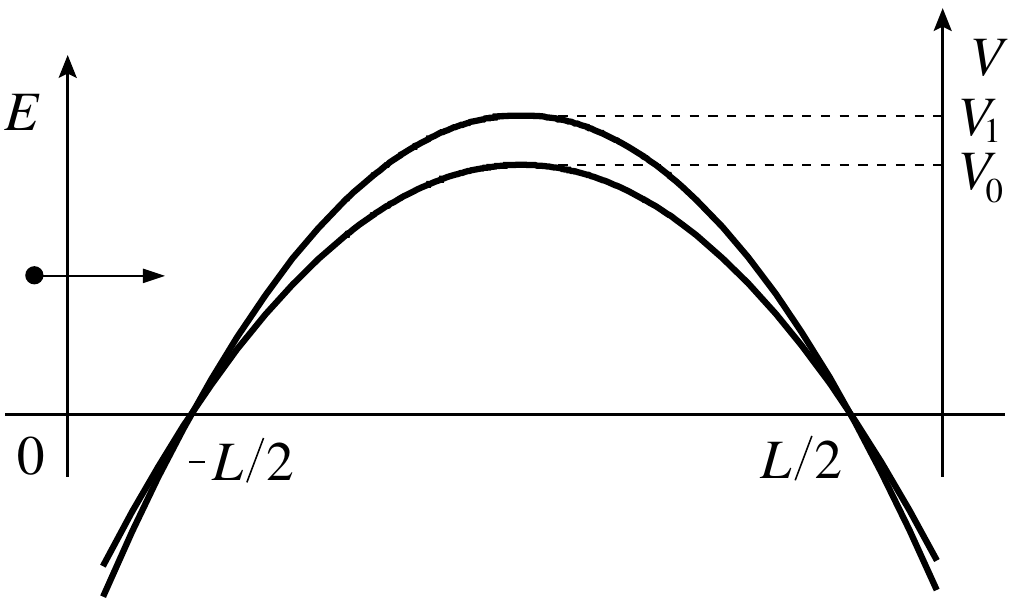}
\caption{QPC modeled by a parabolic potential.}
\label{fig:qpc_qc}
\end{center}
\end{figure}

A more realistic description for the {\it quantum point contact (QPC)} is
achieved by considering a parabolic scattering potential $V_n(x)=V_n-kx^2/2$
where the offset $V_n$ is the QPC barrier height when the dot is in the charge
state $|n\rangle$. Here, we assume a quasi-classical description and consider
the two limits of electrons with energy $E\gg V_n$ resp. $E\ll V_n$, see
Fig.\ \ref{fig:qpc_qc}.

Using the Kemble formula \cite{kemble:35}, we obtain the transmission 
$T_n=1/(1+\exp[-2\pi \sqrt{mL^2/8\hbar^2}(E-V_n)/\sqrt{V_n}])$, where we have
chosen $V(\pm L/2)=0$. For weak coupling $\delta V=V_1-V_0\ll V_0$, we
obtain the shift (we define the energy scale $E_L = \hbar^2/2mL^2$)
\begin{align}
   \delta T=\frac{\pi}{4} \frac{E+V_0}{V_0}\, T_E R_E 
            \frac{\delta V}{\sqrt{E_L V_0}},
\end{align}
which is suppressed exponentially for $E\gg V_0$ and $E\ll V_0$ due to an
exponentially small reflection or transmission.

The change in phase at large energies $E\gg V_0$ is determined by the
transmission phase accumulated in the region $[-L/2,L/2]$; within a
quasi-classical description, this is given by ($\varepsilon \equiv E/V_n$)
\begin{align}
   \theta_n& =\frac{1}{\hbar}\int_{-L/2}^{L/2}dx\,\sqrt{2m(E-V_n(x))}
   \nonumber\\
   &=\frac{1}{2}\sqrt{\frac{V_n}{E_L}} \bigl[ \sqrt{\varepsilon}
   -(\varepsilon - 1) \log(\varepsilon-1)^{1/2}\nonumber\\
   &\qquad\qquad\qquad
   +(\varepsilon - 1) \log(1+\sqrt{\varepsilon}) \bigr].
\end{align}
Expanding this result for small $\delta V$, we obtain the change in phase 
$\delta\theta = \delta \chi$, 
\begin{align}
   \delta\theta\approx 
    -\frac{1}{3}\,\sqrt{\frac{V_{0}}{E}}\, 
     \frac{\delta V}{\sqrt{E_L V_0}}.
\end{align}
Given the exponential suppression of $\delta T$ at large energies $E\gg V_0$,
we find that $\delta T\ll |\delta \theta|= |\delta \chi|$ and for large
transmission we have $|\mathcal{I}S_\mathrm{det,t}|\gg |\mathcal{R}
S_\mathrm{det,r/t}|\gg |\mathcal{I}S_\mathrm{det,r}|$.  In the opposite regime
of small energies $E\ll V_0$, we determine the change in phase (within
quasi-classics) from the phase of the reflection amplitude,
\begin{align}
  \chi_n \approx \frac{2}{\hbar} \int_{-L/2}^{x_0} dx \sqrt{2m(E-V_n(x))},
\end{align}
where the reversal point $x_0<0$ is characterized by $V(x_0)=E$. To leading order 
in $\delta V$ we find that 
\begin{align}
   \delta\chi\approx -\frac{1}{3}\,
   \Bigl(\frac{E}{V_0}\Bigr)^{3/2}\, \frac{\delta V}{\sqrt{E_L V_0}}.
\end{align}
Once more, it follows that $|\delta\theta|,|\delta\chi| \ll \delta T$ due to
the exponential suppression of $T$ and the response functions respect the
order $|\mathcal{R}S_\mathrm{det,r}| \gg |\mathcal{I}S_\mathrm{det,r/t}|\gg
|\mathcal{R}S_\mathrm{det,t}|$. At intermediate energies, the response
functions are of similar magnitude. Summarizing, we find that a scatterer with
large transmission is characterized by the response functions satisfying
$|\mathcal{R}S_\mathrm{det,t}|\gg |\mathcal{I}S_\mathrm{det,r/t}|\gg
|\mathcal{R}S_\mathrm{det,r}|$ while at small transmission
$|\mathcal{R}S_\mathrm{det,r}|\gg |\mathcal{I}S_\mathrm{det,r/t}|\gg
|\mathcal{R}S_\mathrm{det,t}|$.

\end{document}